# Parameterized Regular Expressions and Their Languages

Pablo Barceló  Leonid Libkin  Juan Reutter
U. de Chile  U. of Edinburgh  U. of Edinburgh

**Abstract**

We study regular expressions that use variables, or parameters, which are interpreted as alphabet letters. We consider two classes of languages denoted by such expressions: under the possibility semantics, a word belongs to the language if it is denoted by some regular expression obtained by replacing variables with letters; under the certainly semantics, the word must be denoted by every such expression. Such languages are regular, and we show that they naturally arise in several applications such as querying graph databases and program analysis. As the main contribution of the paper, we provide a complete characterization of the complexity of the main computational problems related to such languages: nonemptiness, universality, containment, membership, as well as the problem of constructing NFAs capturing such languages. We also look at the extension when domains of variables could be arbitrary regular languages, and show that under the certainty semantics, languages remain regular and the complexity of the main computational problems does not change.

## 1 Introduction

In this paper we study parameterized regular expressions like $(0x)^*1(xy)^*$ that combine letters from a finite alphabet $\Sigma$, such as 0 and 1, and variables, such as $x$ and $y$. These variables are interpreted as letters from $\Sigma$. This gives two ways of defining the language of words over $\Sigma$ denoted by a parameterized regular expression $e$. Under the first – possibility – semantics, a word $w \in \Sigma^*$ is in the language $\mathcal{L}_\diamond(e)$ if $w$ is in the language of *some* regular expression $e'$ obtained by substituting alphabet letters for variables. Under the second – certainty – semantics, $w \in \mathcal{L}_\square(e)$ if $w$ is in the language of *all* regular expressions obtained by substituting alphabet letters for variables. For example, if $e = (0x)^*1(xy)^*$, then $01110 \in \mathcal{L}_\diamond(e)$, as witnessed by the substitution $x \mapsto 1, y \mapsto 0$. The word 1 is in $\mathcal{L}_\square(e)$, since the starred subexpressions can be replaced by the empty word. As a more involved example of the certainty semantics, the reader can verify that for $e' = (0|1)^*xy(0|1)^*$, the word 10011 is in $\mathcal{L}_\square(e')$, although no word of length less than 5 can be in $\mathcal{L}_\square(e')$.

These semantics of parameterized regular expressions arise in a variety of applications. In fact, the possibility semantics has already been studied in the case of infinite alphabets [11], with the motivation coming from the study of infinite-state systems with finite control (e.g., software with integer parameters). We, on the other hand, are interested in the classical case of *finite* alphabets, typically considered in connection with formal languages, and both the possibility and the certainty semantics. These are motivated by several applications, in particular in the fields of querying graph-structured data, and static analysis of programs. We now explain these connections.

**Applications in graph databases** Graph databases, that describe both data and its topology, have been actively studied over the past few years in connection with such diverse topics as social networks, biological data, semantic Web and RDF, crime detection and analyzing network traffic; see [2] for a survey. The abstract data model is essentially an edge-labeled graph, with edge labels coming from a finite alphabet. This finite alphabet can contain, for example, types of relationships in a social network or a list of RDF properties. In this setting one concentrates on various types of reachability queries, e.g., queries that ask for the existence of a path between nodes with certain properties so that the label of the path forms a word in a given regular language [3, 6, 7, 9].

As in most data management applications, it is common that some information is missing, typically due to using data that is the result of another query or transformation [1, 4, 8]. For example, in a social network we may have edges $a \xmapsto{x} b$ and $a' \xmapsto{x} b'$, saying that the relationship between $a$ and $b$ is the same as that between $a'$ and $b'$. However, the precise nature of such a relationship is unknown, and this is represented by a variable $x$. Such graphs $G$ whose edges are labeled by letters from $\Sigma$ and variables from a set $\mathcal{W}$ can be viewed as automata over $\Sigma \cup \mathcal{W}$. In checking the existence of paths between nodes, one normally looks for *certain answers* [14], i.e., answers independent of a particular interpretation of variables.

In the case of graph databases such certain answers can be found as follows. Let $a, b$ be two nodes of $G$. One can view $(G, a, b)$ as an automaton, with $a$ as the initial state, and $b$ as the final state; its language, over $\Sigma \cup \mathcal{W}$ is given by some regular expression $e(G, a, b)$. Then we can be certain about the existence of a word $w$ from some language $L$ that is the label of a path from $a$ to $b$ iff $w$ also belongs to $\mathcal{L}_\Box(e(G, a, b))$, i.e., iff $L \cap \mathcal{L}_\Box(e(G, a, b))$ is nonempty. Hence, computing $\mathcal{L}_\Box(e)$ is essential for answering queries over graph databases with missing information.

**Applications in program analysis** That regular expressions with variables appear naturally in program analysis tasks was noticed, for instance, in [18, 19, 20]. One uses the alphabet that consists of symbols related to operations on variables, pointers, or files, e.g., `def` for defining a variable, `use` for using it, `open` for opening a file, or `malloc` for allocating a pointer. A variable then follows: $\text{def}(x)$ means defining variable $x$. While variables and alphabet symbols do not mix freely any more, it is easy to enforce correct syntax with an automaton. An example of a regular condition with parameters is searching for uninitialized variables: $(\neg\text{def}(x))^*\text{use}(x)$.

Expressions like this are evaluated on a graph that serves as an abstraction of a program. One considers two evaluation problems: whether under some evaluation of variables, either some path, or every path between two nodes satisfies it. This amounts to computing $\mathcal{L}_\diamond(e)$ and checking whether all paths, or some path between nodes is in that language. In case of uninitialized variables one would be using 'some path' semantics; the need for the 'all paths' semantics arises when one analyzes locking disciplines or constant folding optimizations [18, 20]. So in this case the language of interest for us is $\mathcal{L}_\diamond(e)$, as one wants to check whether there is an evaluation of variables for which some property of a program is true.

Parameterized regular expressions appeared in other applications as well, e.g., in phase-sequence prediction for dynamic memory allocation [22], or as a compact way to express a family of legal behaviors in hardware verification [5], or as a tool to state regular constraints in constraint satisfaction problems [21].

At the same time, however, very little is known about the basic properties of the languages $\mathcal{L}_\Box(e)$ and $\mathcal{L}_\diamond(e)$. As we mentioned, the $\mathcal{L}_\diamond(e)$ semantics has been studied in the context of *infinite* alphabets [11]. It was shown that it defines languages that can also accepted by non-deterministic register automata of [16]. Language-theoretic issues are quite different over finite and infinite alphabets. In the case of infinite alphabets getting closure properties is nontrivial, and some problems, such universality and containment, are even undecidable [11]. In contrast, in the classical language-theoretic framework of finite alphabets, closure and decidability are guaranteed, and the key question is to pinpoint the precise complexity of the main decision problems.

Thus, our main goal is to determine the exact complexity of the key problems related to languages $\mathcal{L}_\Box(e)$ and $\mathcal{L}_\Diamond(e)$. We consider the standard language-theoretic decision problems, such as membership of a word in the language, language nonemptiness, universality, and containment. Since the languages $\mathcal{L}_\Box(e)$ and $\mathcal{L}_\Diamond(e)$ are regular, we also consider the complexity of constructing NFAs, over the finite alphabet $\Sigma$, that define them.

For all the decision problems, we find complexity classes they belong to. In fact, all the problems end up being complete for various complexity classes, from NLOGSPACE to EXPSPACE. We establish upper bounds on the running time of algorithms for constructing NFAs, and then prove matching lower bounds for the sizes of NFAs representing $\mathcal{L}_\Box(e)$ and $\mathcal{L}_\Diamond(e)$. Finally, we look at some extensions where the range of variables need not be just $\Sigma$. Under the possibility semantics, such languages subsume pattern (and even multi-pattern [15]) languages, but under the certainty semantics they remain regular, and we establish complexity bounds on the main problems.

**Organization** Parameterized regular expressions and their languages are formally defined in Section 2. In Section 3 we define the main problems we study. Complexity of the main decision problems is analyzed in Section 4, and complexity of automata construction in Section 5. In Section 6 we study extensions when domains of variables need not be single letters. In the first 10 pages of the paper we state the main results and provide quick sketches of the proofs; detailed proofs of all the results are in the appendix.

## 2 Preliminaries

Let $\Sigma$ be a finite alphabet, and $\mathcal{V}$ a countably infinite set of variables, disjoint from $\Sigma$. Regular expressions over $\Sigma \cup \mathcal{V}$ will be called *parameterized regular expressions*. Regular expressions, as usual, are built from $\emptyset$, the empty word $\varepsilon$, symbols in $\Sigma$ and $\mathcal{V}$, by operations of concatenation ($\cdot$), union ($|$), and the Kleene star ($*$). Of course each such expression only uses finitely many symbols in $\mathcal{V}$. The size of a regular expression is measured as the total number of symbols needed to write it down (or as the size of its parse tree).

We write $\mathcal{L}(e)$ for the language defined by a regular expression $e$. If $e$ is a parameterized regular expression that uses variables from a finite set $\mathcal{W} \subset \mathcal{V}$, then $\mathcal{L}(e) \subseteq (\Sigma \cup \mathcal{W})^*$. We are interested in languages $\mathcal{L}_\Box(e)$ and $\mathcal{L}_\Diamond(e)$, which are subsets of $\Sigma^*$. To define them, we need the notion of a valuation $\nu$ which is a mapping from $\mathcal{W}$ to $\Sigma$, where $\mathcal{W}$ is the set of variables mentioned in $e$. By $\nu(e)$ we mean the regular expression over $\Sigma$ obtained from $e$ by simultaneously replacing each variable $x \in \mathcal{W}$ by $\nu(x)$. For example, if $e = (0x)^*1(xy)^*$ and $\nu$ is given by $x \mapsto 1, y \mapsto 0$, then $\nu(e) = (01)^*1(10)^*$.

We now formally define the certainty and possibility semantics for parameterized regular expressions.

**Definition 1 (Acceptance)** *Let $e$ be a parameterized regular expression. Then*

$$\begin{aligned} \mathcal{L}_\Box(e) &:= \bigcap \{\mathcal{L}(\nu(e)) \mid \nu \text{ is a valuation for } e\} & \text{(certainty semantics)} \\ \mathcal{L}_\Diamond(e) &:= \bigcup \{\mathcal{L}(\nu(e)) \mid \nu \text{ is a valuation for } e\} & \text{(possibility semantics)} \end{aligned}$$

Since each parameterized regular expression uses finitely many variables, the number of possible valuations is finite as well, and thus both $\mathcal{L}_\Box(e)$ and $\mathcal{L}_\Diamond(e)$ are regular languages over $\Sigma^*$.

The usual connection between regular expressions and automata extends to the parameterized case. Each parameterized regular expression $e$ over $\Sigma \cup \mathcal{W}$, where $\mathcal{W}$ is a finite set of variables in $\mathcal{V}$, can of course be translated, in polynomial time, into an NFA $\mathcal{A}_e$ over $\Sigma \cup \mathcal{W}$ so that $\mathcal{L}(\mathcal{A}_e) = \mathcal{L}(e)$. Such equivalences extend to $\mathcal{L}_\Box$ and $\mathcal{L}_\Diamond$. Namely, for an NFA $\mathcal{A}$ over $\Sigma \cup \mathcal{W}$, and a valuation $\nu : \mathcal{W} \to \Sigma$, define $\nu(\mathcal{A})$ as the NFA over $\Sigma$ that is obtained from $\mathcal{A}$ by replacing each transition of the form $(q, x, q')$ in $\mathcal{A}$ (for $q, q'$ states of $\mathcal{A}$ and $x \in \mathcal{W}$) with the transition $(q, \nu(x), q')$. The following is just an easy observation:

**Lemma 1** *Let $e$ be a parameterized regular expression, and $\mathcal{A}_e$ be an NFA over $\Sigma \cup \mathcal{V}$ such that $\mathcal{L}(\mathcal{A}_e) = \mathcal{L}(e)$. Then, for every valuation $\nu$, we have $\mathcal{L}(\nu(e)) = \mathcal{L}(\nu(\mathcal{A}_e))$.*

Hence, if we define $\mathcal{L}_\Box(\mathcal{A})$ as $\bigcap_\nu \mathcal{L}(\nu(\mathcal{A}))$, and $\mathcal{L}_\Diamond(\mathcal{A})$ as $\bigcup_\nu \mathcal{L}(\nu(\mathcal{A}))$, then the lemma implies that $\mathcal{L}_\Box(e) = \mathcal{L}_\Box(\mathcal{A}_e)$ and $\mathcal{L}_\Diamond(e) = \mathcal{L}_\Diamond(\mathcal{A}_e)$. Since one can go from regular expressions to NFAs in polynomial time, this will allow us to use both automata and regular expressions interchangeably to establish our results.

## 3 Basic Problems

We now describe the main problems we study here. For each problem we shall have two versions, depending on which semantics – $\mathcal{L}_\Box$ or $\mathcal{L}_\Diamond$ is used. So each problem will have a subscript $*$ that can be interpreted as $\Box$ or $\Diamond$.

We start with decision problems:

NONEMPTINESS$_*$  Given a parameterized regular expression $e$, is $\mathcal{L}_*(e) \neq \emptyset$?

MEMBERSHIP$_*$  Given a parameterized regular expression $e$ and a word $w \in \Sigma^*$, is $w \in \mathcal{L}_*(e)$?

UNIVERSALITY$_*$  Given a parameterized regular expression $e$, is $\mathcal{L}_*(e) = \Sigma^*$?

CONTAINMENT$_*$  Given parameterized regular expressions $e_1$ and $e_2$, is $\mathcal{L}_*(e_1) \subseteq \mathcal{L}_*(e_2)$?

A special version of nonemptiness is the problem of intersection with a regular language (used in the database querying example in the introduction):

NONEMPTYINTREG$_*$  Given a parameterized regular expression $e$, and a regular expression $e'$ over $\Sigma$, is $\mathcal{L}(e') \cap \mathcal{L}_*(e) \neq \emptyset$?

The last problem is computational rather than a decision problem:

CONSTRUCTNFA$_*$  Given a parameterized regular expression $e$, construct an NFA $\mathcal{A}$ over $\Sigma$ such that $\mathcal{L}_*(e) = \mathcal{L}(\mathcal{A})$.

# 4 Decision problems

In this section we consider the five decision problems – nonemptiness, membership, universality, containment, as well as nonemptiness of intersection with a regular language – and provide precise complexity for them.

We shall also consider two restrictions on regular expressions; these will indicate when the problems are inherently very hard or when their complexity can be lowered in some cases. One source of complexity is the repetition of variables in expressions like $(0x)^*1(xy)^*$. When no variable appears more than once in a parameterized regular expression, we call it *simple*. Another source of complexity is infinite languages, so we consider a restriction to expressions of *star-height* 0, in which no Kleene star is used: these denote finite languages, and each finite language is denoted by such an expression.

## 4.1 Nonemptiness

The problem NONEMPTINESS$_\diamond$ has a trivial solution, since $\mathcal{L}_\diamond(e) \neq \emptyset$ for every parameterized regular expression $e$ (except $e = \emptyset$). So we study this problem for the certainty semantics; for the possibility semantics, we look at the related problem NONEMPTINESS-AUTOMATA$_\diamond$, which, for a given NFA $\mathcal{A}$ over $\Sigma \cup \mathcal{V}$ asks whether $\mathcal{L}_\diamond(\mathcal{A}) \neq \emptyset$.

**Theorem 1**
- *The problem* NONEMPTINESS$_\square$ *is* EXPSPACE-*complete.*
- *The problem* NONEMPTINESS-AUTOMATA$_\diamond$ *is* NLOGSPACE-*complete.*

The result for the possibility semantics is by a standard reachability argument. Note that the bound is the same here as in the case of infinite alphabets studied in [11]. To see the upper bound for NONEMPTINESS$_\square$, note that there are exponentially many valuations $\nu$, and each automaton $\nu(\mathcal{A}_e)$ is of polynomial size, so we can use the standard algorithm for checking nonemptiness of the intersection of a family of regular languages which can be solved in polynomial space in terms of the size of its input; since the input to this problem is of exponential size in terms of the original input, the EXPSPACE bound follows. The hardness is by a generic (Turing machine) reduction; the proof is in the appendix.

In the proof we use the following property of the certainty semantics, which shows a striking difference with the case of standard regular expressions:

**Lemma 2** *Given a set $e_1, \ldots, e_k$ of parameterized expressions of size at most $n \geq k$, it is possible to build, in time $O(k \cdot n)$ an expression $e'$ such that $\mathcal{L}_\square(e')$ is empty if and only if $\mathcal{L}_\square(e_1) \cap \cdots \cap \mathcal{L}_\square(e_k)$ is empty.*

The reason the case of the $\mathcal{L}_\square(e)$ semantics is so different from the usual semantics of regular languages is as follows. It is well known that checking whether the intersection of the languages defined by a finite set $S$ of regular expressions is nonempty is PSPACE-complete [17], and hence under widely held complexity-theoretic assumptions no regular expression $r$ can be constructed in polynomial time from $S$ such that $\mathcal{L}(r)$ is nonempty if and only if $\bigcap_{s \in S} \mathcal{L}(s)$ is nonempty. Lemma 3, on the other hand, says that such a construction is possible for parameterized regular expressions under the certainty semantics.

The generic reduction used in the proof of EXPSPACE-hardness of NONEMPTINESS$_\square$ also provides lower bounds on the minimal sizes of words in languages $\mathcal{L}_\square(e)$ (note that the language $\mathcal{L}_\diamond(e)$ always contains a word of the size linear in the size of $e$).

**Corollary 1** *There exists a polynomial $p : \mathbb{N} \to \mathbb{N}$ and a sequence of parameterized regular expressions $\{e_n\}_{n \in \mathbb{N}}$ such that each $e_n$ is of size at most $p(n)$, and every word in the language $\mathcal{L}_\Box(e_n)$ has size at least $2^{2^n}$.*

The construction is somewhat involved, but it is easy to see the single-exponential bound (which was hinted at in the first paragraph of the introduction, and which was in fact used in connection with querying incomplete graph data in [4]). For each $n$, consider an expression $e_n = (0|1)^* x_1 \ldots x_n (0|1)^*$. If a word $w$ is in $\mathcal{L}_\Box(e_n)$, then $w$ must contain every word in $\{0,1\}^n$ as a subword, which implies that its length must be at least $2^n + 1$.

We can also show that the use of Kleene star has a huge impact on complexity, which is not at the same time affected by variable repetitions.

**Proposition 1** *The problem $\mathrm{NONEMPTINESS}_\Box$ remains EXPSPACE-hard over the class of simple regular expressions, but it is $\Sigma_2^p$-complete over the class of expressions of star-height 0.*

## 4.2 Membership

It is easy to see that $\mathrm{MEMBERSHIP}_\Box$ can be solved in coNP, and $\mathrm{MEMBERSHIP}_\Diamond$ in NP: one just guesses a valuation witnessing $w \in \mathcal{L}(v(e))$ or $w \notin \mathcal{L}(v(e))$. These bounds turn out to be tight.

**Theorem 2**
- *The problem $\mathrm{MEMBERSHIP}_\Box$ is coNP-complete.*
- *The problem $\mathrm{MEMBERSHIP}_\Diamond$ is NP-complete.*

Note that for the case of the possibility semantics, the bound is the same as for languages over the infinite alphabets [11] (for all problems other than nonemptiness and membership, the bounds will be different). The hardness proof in [11] relies on the infinite size of the alphabet, but one can find an alternative proof that uses only finitely many symbols. Both proofs are by variations of 3-SAT or its complement; the reductions are somewhat involved and are presented in the appendix.

The restrictions to expressions without repetitions, or to finite languages, by themselves do not lower the complexity, but together they make it polynomial.

**Proposition 2** *The complexity of the membership problem remains as in Theorem 2 over the classes of simple expressions, and expressions of star-height 0. Over the class of simple expressions of star-height 0, $\mathrm{MEMBERSHIP}_\Diamond$ can be solved in time $O(nm \cdot \log^2 n)$, where $n$ is the size of the expression and $m$ is the size of the word.*

The $\log^2 n$ factor appears due to the complexity of the algorithm for converting regular expressions into $\varepsilon$-free NFAs [13]; it appears as one of the steps of the algorithm, which is described in the appendix.

**Membership for fixed words** We next consider a variation of the membership problem: $\mathrm{MEMBERSHIP}_*(w)$ asks, for a parameterized regular expression $e$, whether $w \in \mathcal{L}_*(e)$. In other words, $w$ is fixed. It turns out that for the $\Diamond$-semantics, this version is efficiently solvable, but for the $\Box$-semantics, it remains intractable unless restricted to simple expressions.

**Theorem 3**
- *There is a word $w \in \Sigma^*$ such that the problem $\mathrm{MEMBERSHIP}_\Box(w)$ is coNP-hard (even over the class of expressions of star-height 0).*

- *For each word $w \in \Sigma^*$, the problem* MEMBERSHIP$_\Box(w)$ *is solvable in $O(n)$ time, if restricted to the class of simple expressions.*

- *For each word $w \in \Sigma^*$, the problem* MEMBERSHIP$_\Diamond(w)$ *is solvable in time $O(n \log^2 n)$.*

## 4.3 Universality

Somewhat curiously, the universality problem is more complex for the possibility semantics $\mathcal{L}_\Diamond$. Indeed, consider a parameterized expression $e$ over $\Sigma$, with variables in $\mathcal{W}$. For the certainty semantics, it suffices to guess a word $w$ and a valuation $\nu : \mathcal{W} \to \Sigma$ such that $w \notin \mathcal{L}(\nu(e))$. This gives a PSPACE upper bound for this problem, which is the best that we can do, as the universality problem is PSPACE-hard even for complete regular expressions. On the other hand, when solving this problem for the possibility semantics, one can expect that all possible valuations for $e$ will need to be analyzed, which increases the complexity by one exponential. (In fact, when one moves to infinite alphabets, this problem becomes undecidable [11]). The lower bound proof, which is again by a generic reduction, is in the appendix.

**Theorem 4**  
- *The problem* UNIVERSALITY$_\Box$ *is* PSPACE-*complete.*
- *The problem* UNIVERSALITY$_\Diamond$ *is* EXPSPACE-*complete.*

Similarly to the nonemptiness problem, the EXPSPACE bound for UNIVERSALITY$_\Diamond$ is quite resilient, as it holds even if for simple expressions (note that it makes no sense to study expressions of star-height 0, as they denote finite languages and thus cannot be universal).

**Proposition 3** *The problem* UNIVERSALITY$_\Diamond$ *remains* EXPSPACE-*hard over the class of simple parameterized regular expressions.*

## 4.4 Containment

The bounds for the containment problem are easily obtained from the fact that both nonemptiness and universality can be cast as its versions. Since $\Sigma^* \subseteq \mathcal{L}_\Diamond(e)$ iff UNIVERSALITY$_\Diamond(e)$ is true, and $\mathcal{L}_\Box(e) \subseteq \emptyset$ iff NONEMPTINESS$_\Box(e)$ is false, we get EXPSPACE lower bounds for both containment problems. The matching upper bounds are by straightforward enumeration of valuations. Hence, we get:

**Theorem 5** *Both* CONTAINMENT$_\Box$ *and* CONTAINMENT$_\Diamond$ *are* EXPSPACE-*complete.*

**Containment with one fixed expression** We look at two variations of the containment problem, when one of the expressions is fixed: CONTAINMENT$_*(e_1, \cdot)$ asks, for a parameterized regular expression $e_2$, whether $\mathcal{L}_*(e_1) \subseteq \mathcal{L}_*(e_2)$; and CONTAINMENT$_*(\cdot, e_2)$ is defined similarly. Theorem 5 show that CONTAINMENT$_\Box(\cdot, e_2)$ and CONTAINMENT$_\Diamond(e_1, \cdot)$ remain EXPSPACE-complete. For the other two versions of the problem, the proposition below (proved in the appendix) shows that the complexity is lowered by at least one exponential.

**Proposition 4**  
- CONTAINMENT$_\Box(e_1, \cdot)$ *is* PSPACE-*complete.*
- CONTAINMENT$_\Diamond(\cdot, e_2)$ *is* coNP-*complete.*

# 5 Computing automata

In this section, we first provide upper bounds for algorithms for building NFAs over $\Sigma$ capturing $\mathcal{L}_\Diamond(e)$ and $\mathcal{L}_\Box(e)$, and then prove their optimality, by showing matching lower bounds on the sizes of such NFAs. Recall that we are dealing with the problem ConstructNFA$_*$: given a parameterized regular expression $e$, construct an NFA $\mathcal{A}$ over $\Sigma$ so that $\mathcal{L}(\mathcal{A}) = \mathcal{L}_*(e)$.

**Proposition 5** *The problem* ConstructNFA$_\Diamond$ *can be solved in single-exponential time, and the problem* ConstructNFA$_\Box$ *can be solved in double-exponential time.*

These bounds are achieved by using naive algorithms for constructing automata: namely, one converts a parameterized regular expression $e$ over variables in a finite set $\mathcal{W}$ into an automaton $\mathcal{A}_e$, and then for $|\Sigma|^{|\mathcal{W}|}$ valuations $\nu$ computes the automata $\nu(\mathcal{A}_e)$. This takes exponential time. To obtain an NFA for $\mathcal{L}_\Diamond(e)$ one simply combines them with a nondeterministic choice; for $\mathcal{L}_\Box(e)$ one takes the product of them, resulting in double-exponential time.

We now show that these complexities are unavoidable, as the smallest NFAs capturing $\mathcal{L}_\Diamond(e)$ or $\mathcal{L}_\Box(e)$ can be of single or double-exponential size, respectively. We say that the *sizes of minimal NFAs for $\mathcal{L}_*$ are necessarily exponential (resp., double-exponential)* if there exists a family $\{e_n\}_{n \in \mathbb{N}}$ of parameterized regular expressions such that:

- the size of each $e_n$ is $O(n)$, and
- every NFA $\mathcal{A}$ satisfying $\mathcal{L}(\mathcal{A}) = \mathcal{L}_*(e_n)$ has at least $2^n$ (resp., $2^{2^n}$) states.

**Theorem 6** *The sizes of minimal NFAs are necessarily double-exponential for $\mathcal{L}_\Box$, and necessarily exponential for $\mathcal{L}_\Diamond$.*

*Proof sketch:* We begin with the double exponential bound for $\mathcal{L}_\Box$. For each $n \in \mathbb{N}$, let $e_n$ be the following parameterized regular expression over alphabet $\Sigma = \{0, 1\}$ and variables $x_1, \ldots, x_{n+1}$:

$$e_n = ((0 \mid 1)^{n+1})^* \cdot x_1 \cdots x_n \cdot x_{n+1} \cdot ((0 \mid 1)^{n+1})^*.$$

Notice that each $e_n$ uses $n + 1$ variables, and is of linear size in $n$. In order to show that every NFA deciding $\mathcal{L}_\Box(e_n)$ has $2^{2^n}$ states, we use the following result from [10]: if $L \subset \Sigma^*$ is a regular language, and there exists a set of pairs $P = \{(u_i, v_i) \mid 1 \leq i \leq m\} \subseteq \Sigma^* \times \Sigma^*$ such that (1) $u_i v_i \in L$, for every $1 \leq i \leq m$, and (2) $u_j v_i \notin L$, for every $1 \leq i, j \leq m$ and $i \neq j$, then every NFA accepting $L$ has at least $m$ states.

Given a collection $S$ of words over $\{0, 1\}$, let $w_S$ denote the concatenation, in lexicographical order, of all the words that belong to $S$, and let $w_{\bar{S},n}$ denote the concatenation of all words in $\{0,1\}^{n+1}$ that are not in $S$.

Then, define a set of pairs $P_n = \{(w_S, w_{\bar{S},n}) \mid S \subset \{0,1\}^{n+1} \text{ and } |S| = 2^n\}$. Since there are $2^{n+1}$ binary words of length $n + 1$, there are $\binom{2^{n+1}}{2^n}$ different subsets of $\{0,1\}^{n+1}$ of size $2^n$, and thus $P_n$ contains $\binom{2^{n+1}}{2^n} \geq 2^{2^n}$ pairs. Moreover, one can show (details in the appendix) that $\mathcal{L}_\Box(e_n)$ and $P_n$ satisfy properties (1) and (2) above, which proves the double exponential lower bound.

To show the exponential lower bound for $\mathcal{L}_\Diamond$, define $e_n = (x_1 \cdots x_n)^*$, and let $P_n = \{(w, w) \mid w \in \{0,1\}^n\}$. Clearly, $P_n$ contains $2^n$ pairs. All that is left to do is to show that $\mathcal{L}_\Diamond(e_n)$ and $P_n$ satisfy properties (1) and (2) above. Details are in the appendix. □

Note that the bounds of Theorem 6 apply to simple regular expressions.

The table in Fig. 1 summarizes the main results in Sections 4 and 5.

| Problem \ Semantics | Certainty □ | Possibility ◇ |
|---|---|---|
| NONEMPTINESS | EXPSPACE-complete | NLOGSPACE-complete (for automata) |
| MEMBERSHIP | coNP-complete | NP-complete |
| CONTAINMENT | EXPSPACE-complete | EXPSPACE-complete |
| UNIVERSALITY | PSPACE-complete | EXPSPACE-complete |
| NONEMPTYINTREG | EXPSPACE-complete | NP-complete |
| CONSTRUCTNFA | double-exponential | single-exponential |

Figure 1: Summary of complexity results

# 6 Extending domains of variables

So far we assumed that variables take values in $\Sigma$: our valuations were partial maps $\nu : \mathcal{V} \to \Sigma$. We now consider a more general case when the range of each variable is a regular subset of $\Sigma^*$.

Let $e$ be a parameterized regular expression with variables $x_1, \ldots, x_n$, and let $L_1, \ldots, L_n \subseteq \Sigma^*$ be nonempty regular languages. We shall write $\bar{L}$ for $(L_1, \ldots, L_n)$. A valuation in $\bar{L}$ is a map $\nu : \bar{x} \to \bar{L}$ such that $\nu(x_i) \in L_i$ for each $i \leq n$. Under such a valuation, each parameterized regular expression $e$ is mapped into a usual regular expression $\nu(e)$ over $\Sigma$, in which each variable $x_i$ is replaced by the word $\nu(x_i)$. Hence we can still define

$$\begin{aligned}\mathcal{L}_\square(e; \bar{L}) &= \bigcap\{\mathcal{L}(\nu(e)) \mid \nu \text{ is a valuation over } \bar{L}\} \\ \mathcal{L}_\diamond(e; \bar{L}) &= \bigcup\{\mathcal{L}(\nu(e)) \mid \nu \text{ is a valuation over } \bar{L}\}\end{aligned}$$

According to this notation, $\mathcal{L}_\square(e) = \mathcal{L}_\square(e; (\Sigma, \ldots, \Sigma))$, and likewise for $\mathcal{L}_\diamond$.

Note however that intersections and unions are now infinite, if some of the languages $L_i$'s are infinite, so we cannot conclude, as before, that we deal with regular languages. And indeed they are not: for example, $\mathcal{L}_\diamond(xx; \Sigma^*)$ is the set of square words, and thus not regular.

We now consider two cases. If each $L_i$ is a finite language, we show that all the complexity results in Fig. 1 remain true. Then we look at the case of arbitrary regular $L_i$'s. Languages $\mathcal{L}_\diamond(e; \bar{L})$ need not be regular anymore, but languages $\mathcal{L}_\square(e; \bar{L})$ still are, and we prove that the complexity bounds from the certainty column of Fig. 1 remain true. For complexity results, we assume that in the input $(e; \bar{L})$, each domain $L_i$ is given either as a regular expression or an NFA over $\Sigma$.

## 6.1 The case of finite domains

If all domain languages $L_i$'s are finite, all the lower bounds apply (they were shown when each $L_i = \Sigma$). For upper bounds, note that each finite $L_i$ contains at most exponentially many words in the size of either a regular expression or an NFA that gives it, and each such word is polynomial size. Thus, the number of valuations is at most exponential in the size of the input, and each valuation can be represented in polynomial time. The following is then straightforward.

**Proposition 6** *If domains $L_i$'s of all variables are finite nonempty subsets of $\Sigma^*$, then both $\mathcal{L}_\square(e; \bar{L})$ and $\mathcal{L}_\diamond(e; \bar{L})$ are regular languages, and all the complexity bounds on the problems related to them are exactly the same as stated in Fig. 1.*

## 6.2 The case of infinite domains

We have already seen that if just one of the domains is infinite, then $\mathcal{L}_\diamond(e; \bar{L})$ need not be regular (the $\mathcal{L}_\diamond(xx; \Sigma^*)$ example). Somewhat surprisingly, however, in the case of the certainty semantics, we recover not only regularity but also all the complexity bounds.

**Theorem 7** *For each parameterized regular expression $e$ using variables $x_1, \ldots, x_n$ and for each an n-tuple $\bar{L}$ of regular languages over $\Sigma$, the language $\mathcal{L}_\square(e; \bar{L}) \subseteq \Sigma^*$ is regular. Moreover, the complexity bounds are exactly the same as in the $\square$ column of the table in Fig. 1.*

*Proof sketch:* We only need to be concerned about regularity of $\mathcal{L}_\square(e; \bar{L})$ and upper complexity bounds, as the proofs of lower bounds apply for the case when all $L_i = \Sigma$. For this, it suffices to prove that there is a finite set $U$ of NFAs so that $\mathcal{L}_\square(e; \bar{L}) = \bigcap_{\mathcal{A} \in U} \mathcal{L}(\mathcal{A})$. Moreover, it follows from analyzing the proofs of upper complexity bounds, that the complexity results will remain the same if the following can be shown about the set $U$:

- its size is at most exponential in the size of the input;
- checking whether $\mathcal{A} \in U$ can be done in time polynomial in the size of $\mathcal{A}$;
- each $\mathcal{A} \in U$ is of size polynomial in the size of the input $(e; \bar{L})$.

To show these, take $\mathcal{A}_e$ and from it construct a reduced automaton $\mathcal{A}'_e$ in which all transitions $(q, x_i, q')$ are eliminated whenever $L_i$ is infinite. We then show that $\mathcal{L}_\square(\mathcal{A}_e; \bar{L}) = \mathcal{L}_\square(\mathcal{A}'_e; \bar{L})$ (the definition of $\mathcal{L}_\square$ extends naturally from regular expressions to automata for arbitrary domains). This observation generates a finite set $U$ of NFAs which result from applying valuations with finite codomains to $\mathcal{A}'_e$. We prove in the appendix that these automata satisfy the required properties.

## 7 Future work

For most bounds (except universality and containment), the complexity under the possibility semantics is reasonable, while for the certainty semantics it is quite high (i.e., double-exponential in practice). At the same time, the concept of $\mathcal{L}_\square(e)$ captures many query answering scenarios over graph databases with incomplete information [4]. One of the future directions of this work is to devise better algorithms for problems related to the certainty semantics under restrictions arising in the context of querying graph databases.

Another line of work has to do with closure properties: we know that results of Boolean operations on languages $\mathcal{L}_\square(e)$ and $\mathcal{L}_\diamond(e)$ are regular and can be represented by NFAs; the bounds on sizes of such NFAs follow from the results shown here. However, it is conceivable that such NFAs can be succinctly represented by parameterized regular expressions. To be concrete, one can easily derive from results in Section 5 that there is a doubly-exponential size NFA $\mathcal{A}$ so that $\mathcal{L}(\mathcal{A}) = \mathcal{L}_\square(e_1) \cap \mathcal{L}_\square(e_2)$, and that this bound is optimal. However, it leaves open a possibility that there is a much more succinct parameterized regular expression $e$ so that $\mathcal{L}_\square(e) = \mathcal{L}_\square(e_1) \cap \mathcal{L}_\square(e_2)$; in fact, nothing that we have shown contradicts the existence of a polynomial-size expression with this property. We plan to study bounds on such regular expressions in the future.

# 8 APPENDIX: COMPLETE PROOFS

**Proof of Theorem 1:**

(Part 1) We begin with the upper bound. Let $e$ be a parameterized regular expression, and as usual assume that $\mathcal{W}$ is the set of variables mentioned in $e$. By definition, $\mathcal{L}_\Box$ is defined as the intersection of each $L(v(e))$, for all possible valuations $\nu : \mathcal{W} \to \Sigma$ for $e$. Clearly, the total number of those valuations is $|\Sigma|^{|\mathcal{W}|}$. Thus, since there are only exponentially many valuations, an EXPSPACE algorithm can guess, symbol by symbol, a word $w \in \mathcal{L}_\Box(e)$, while checking, in parallel, that $w$ belongs to each $L(\nu(e))$, for every such valuation $\nu : \mathcal{W} \to \Sigma$.

The proof for the lower bound relies heavily on the following lemma:

**Lemma 3** *Given a set $e_1, \ldots, e_k$ of parameterized expressions over an alphabet $\Sigma$ that contains at least two symbols, it is possible to build, in polynomial time with respect to $e_1, \ldots, e_k$, an expression $e'$ such that $\mathcal{L}_\Box(e')$ is empty if and only if $\mathcal{L}_\Box(e_1) \cap \cdots \cap \mathcal{L}_\Box(e_k)$ is empty.*

*Proof:* Let $e_1, \ldots, e_k$ as stated in the Lemma, and let $a, b \in \Sigma$. We use $(\Sigma - a)$ as a shorthand for the expression whose language is the union of every symbol in $\Sigma$ different from $a$, and define $A^i = [(\Sigma - a)^* \cdot a \cdot (\Sigma - a)^*]^i$, for $1 \leq i \leq k - 1$. Finally, let $x_1, \ldots, x_{k-1}$ be fresh variables. We define $e'$ as

$$(\Sigma - a)^* \cdot x_1 \cdot (\Sigma - a)^* \cdot x_2 \cdot (\Sigma - a)^* \cdots x_{k-1} \cdot (\Sigma - a)^* \cdot$$
$$(ba^k b \cdot e_1 \mid b \cdot A^1 \cdot ba^k b \cdot e_2 \mid b \cdot A^2 \cdot ba^k b \cdot e_3 \mid \cdots \mid b \cdot A^{k-1} \cdot ba^k b \cdot e_k).$$

To prove the lemma, consider first that a word $w$ that belongs to $\mathcal{L}_\Box(e_1) \cap \cdots \cap \mathcal{L}_\Box(e_k)$. Then, it can be proved that the word $(\bar{c}^k ab)^{k-1} ba^k bw$ belongs to $\mathcal{L}_\Box(e')$ where $\bar{c}$ is the concatenation (say, in lexicographical order) of all the symbols in $\Sigma$ different from $a$. On the other hand, assume that a word $w$ belongs to $\mathcal{L}_\Box(e')$. We need the following claim:

**Claim 1** *The word $w$ can be decomposed into $u \cdot ba^k b \cdot v$, where $u$ contains exactly $k-1$ appearances of $a$.*

*Proof:* By the inspection of $e'$, we conclude that $w$ must contain the substring $ba^k b$. Now, let $w = u \cdot ba^k b \cdot v$, where $u$ does not contain substring $ba^k b$. Assume first that $u$ contains less than $k - 1$ appearances of the symbol $a$. Then, consider a valuation $\nu$ that maps each variable in $e'$ to the symbol $a$. Since $\nu(e')$ is of form

$$((\Sigma - a)^* a)^{k-1} (\Sigma - a)^* \cdot (ba^k b \cdot \nu(e_1) \mid b \cdot A^1 \cdot ba^k b \cdot \nu(e_2) \mid b \cdot A^2 \cdot ba^k b \cdot \nu(e_3) \mid \cdots \mid b \cdot A^{k-1} \cdot ba^k b \cdot \nu(e_k)).$$

we conclude that the language of $\nu(e')$ cannot contain any word that starts with $uba^k b$, since all the words in $\mathcal{L}(\nu(e'))$ must start with a prefix in the language

$$((\Sigma - a)^* a)^{k-1} (\Sigma - a)^* \cdot b.$$

Next, assume that $u$ contains more than $k - 1$ appearances of the symbol $a$, an consider a valuation $\nu'$ that maps each variable in $e'$ to the symbol $b$. Then again, $\nu'(e')$ is of form

$$((\Sigma - a)^* b)^{k-1} (\Sigma - a)^* (ba^k b \cdot \nu'(e_1) \mid b \cdot A^1 \cdot ba^k b \cdot \nu'(e_2) \mid b \cdot A^2 \cdot ba^k b \cdot \nu'(e_3) \mid \cdots \mid b \cdot A^{k-1} \cdot ba^k b \cdot \nu'(e_k)).$$

Then, notice that any word in $\mathcal{L}(\nu'(e'))$ is such that the symbol $a$ cannot appear more than $k-1$ times before the substring $ba^kb$. We conclude that $\mathcal{L}(\nu'(e'))$ cannot contain a word starting with $u \cdot ba^kb \cdot v$. □

Using this Claim, It is now straightforward to show that, if a valuation $\nu$ assigns the symbol $a$ to exactly $j$ variables in $\{x_1, \ldots, x_{k-1}\}$ ($1 \leq j \leq k-1$), then $v$ must belong to $\mathcal{L}_\square(e_{k-j-1})$. This shows that $v$ belongs to $\mathcal{L}_\square(e_1) \cap \cdots \cap \mathcal{L}_\square(e_k)$, which finishes the proof of the Lemma. □.

Next we continue with the proof of the theorem, specifically, we show the EXPSPACE lower bound for the problem NONEMPTINESS$_\square$. The proof is by a reduction from the acceptance problem of Turing machines. Let $L$ be a language that belongs to EXPSPACE, and let $\mathcal{M}$ be a Turing machine that decides $L$ in EXPSPACE. Given an input $\bar{a} = a_0, \ldots, a_{k-1}$, we construct in polynomial time with respect to $\mathcal{M}$ and $\bar{a}$ a parameterized regular expression $e_{\mathcal{M}, \bar{a}}$ such that $\mathcal{L}_\square(e_{\mathcal{M}, \bar{a}}) \neq \emptyset$ if and only if $\mathcal{M}$ accepts $\bar{a}$.

Assume that $\mathcal{M} = \{Q_\mathcal{M}, \Sigma_\mathcal{M}, \Sigma_\mathcal{M} \cup \{B\}, q_0, \{q_m\}, \delta_\mathcal{M}\}$; that is, the states of $\mathcal{M}$ are $Q_\mathcal{M} = \{q_0, \ldots, q_m\}$, the initial state is $q_0$, the alphabet of the input $\mathcal{M}$ is $\Sigma_\mathcal{M}$, and the alphabet of $\mathcal{M}$ is $\Gamma_\mathcal{M} = \Sigma_\mathcal{M} \cup \{B\}$, that is, $\mathcal{M}$ uses an additional *blank* symbol $B$; and that the set of transitions of $\mathcal{M}$ is $\delta_\mathcal{M}$. Without loss of generality, we assume that $\mathcal{M}$ has only one tape, starts with the input copied on the first $|\bar{a}|$ cells of its tape, that $\mathcal{M}$ has only one final state $q_m$, and that no transition is defined for that state. Moreover, it will be easier for us to assume that the machine always end after an odd number of steps (although the reduction can be improved to work without this assumption). Since $\mathcal{M}$ decides $L$ in EXPSPACE, there is a polynomial $S()$ such that, for every input $\bar{a}$ over $\Sigma_\mathcal{M}$, $\mathcal{M}$ decides $\bar{a}$ using space of order $2^{S(|\bar{a}|)}$. Assume for notation convenience that $S(|\bar{a}|) = n$.

Due to Lemma 3, it suffices to construct in polynomial time a set $E$ containing a polynomial number of parameterized regular expressions, such that $\bigcap_{e \in E} \mathcal{L}_\square(e)$ is empty of and only if $\mathcal{M}$ accepts on input $\bar{a}$. But before we describe the set $E$, we need some notation. We use the shorthand $[i]$ to denote the binary representation of the number $i$ as a string of $n$ characters. For example, $[0]$ corresponds to the word $0^n$, and $[2]$ corresponds to the word $0^{n-2}10$. Roughly speaking, any word accepted by all the expressions in $E$ should represent a valid run of the Turing machine. In order to do so, $E$ is constructed so that all words that belong to $\bigcap_{e \in E} \mathcal{L}_\square(e)$ represent a sequence

$$[even] \cdot L_0 \cdot [even] \cdot [odd] \cdot L_1 \cdot [odd] \cdot [even] \cdot L_2 \cdot [even] \cdots,$$

where each of the $L$'s represent a configuration of $\mathcal{M}$, coded as

$$(action \cdot stateb \cdot [0] \cdot statea \cdot action \cdot stateb \cdot [1] \cdot statea \cdots action \cdot stateb \cdot [2^n - 1] \cdot statea)^*.$$

Each construct $action \cdot state \cdot [i] \cdot state$ represents the content of the $i$-th cell of the Turing machine before and after one given point of the computation, plus the action that was performed in that cell in the given step step. More precisely, the word *action* is a three bit string that codes either nothing, read/write, or advance head, *stateb* is a binary word coding $\Sigma_\mathcal{M} \cup \Sigma_\mathcal{M} \times Q_\mathcal{M}$, representing the content of the $i$-th cell before a given step in the computation of $\mathcal{M}$ (if *stateb* codes a pair in $\Sigma_\mathcal{M} \times Q_\mathcal{M}$ this represents that $\mathcal{M}$ was pointing at that cell), $[i]$ is the binary representation of the number $i$ ($0 \leq i \leq 2^n - 1$) that represents the cell's position, and *statea* is a binary word coding $\Sigma_\mathcal{M} \cup \Sigma_\mathcal{M} \times Q_\mathcal{M}$ that represents the content of the $i$-th cell after the computation.

Thus, intuitively, each of this words can be seen as a sequence of descriptions of $M$, each description consisting of a sequence of tuples that encode, for each cell, the construct *action*, *state* before the action, *position* and *state* after the action. The idea behind the reduction is that $E$ should only accept those words representing a valid computation, that ends in a final state. The rest of the reduction is devoted to construct such set $E$. We divide the set $E$ into sets $E_{form}$, $E_i$, $E_f$, $E_{order}$ and $E_{state}$. But let us first show our coding scheme.

Let $p = \log(\Gamma + |\Gamma||Q|)$. We code each content of a cell (that is, strings corresponding to *statea* and *stateb* coding $\Gamma \cup \Gamma \times Q_\mathcal{M}$ as a $p$-bit string. Moreover, we code the actions as 3-bit strings. For simplicity, we denote this strings by $[even]$, $[odd]$, $[nothing]$, $[read]$ and $[head]$; where $[even] = 000$, $[odd] = 001$, $[nothing] = 100$, $[read] = 101$ and $[head] = 111$. Finally, we use $[action]$ as a shorthand for $[nothing] \mid [read] \mid [head]$ (or, equivalently, $100 \mid 101 \mid 111$).

We now define the expressions in $E$. First, $E_{form}$ contains the expression

$$\big([even][action](0 \mid 1)^p(0 \mid 1)^n(0 \mid 1)^p[even][odd][action](0 \mid 1)^p(0 \mid 1)^n(0 \mid 1)^p[odd]\big)^*$$

The intuition behind $E_{form}$ is that it defines that each word in $\bigcap_{e \in E} \mathcal{L}_\square(e)$ is of the form that was explained in the previous paragraphs.

Next, we define $E_{order}$, that intuitively forces each of the description of the $L$'s in $\bigcap_{e \in E} \mathcal{L}_\square(e)$ to be of form

$$(action \cdot stateb \cdot [0] \cdot statea \cdot action \cdot stateb \cdot [1] \cdot statea \cdots action \cdot stateb \cdot [2^n - 1] \cdot statea)^*.$$

That is, the slots used to code the position of the cell in each description have to be arranged in numerical order. We split $E_{order}$ into $E_{order}^{even}$ and $E_{order}^{odd}$. The sets are as follows: $E_{order}^{even}$ contains the expression

$$\bigg(\big([even] \mid [odd][action](0 \mid 1)^p[0](0 \mid 1)^p[even] + [odd]\big) \cdot$$

$$\big([even] \mid [odd][action](0 \mid 1)^p(0 \mid 1)^n(0 \mid 1)^p[even] + [odd]\big)^*\bigg)^*$$

stating, intuitively, that each portion of the words belonging to $E_{order}$ start with a $[0]$ in the slot corresponding to the cell position. In addition, for each $1 \leq m \leq n-1$, $E_{order}^{even}$ contains the expression

$$\big([even] \mid [odd]$$

$$\bigg(\big([action](0 \mid 1)^p(0 \mid 1)^{n-m-1}0(0 \mid 1)^{m-1}0(0 \mid 1)^p[action](0 \mid 1)^p(0 \mid 1)^{n-m-1}0(0 \mid 1)^{m-1}1(0 \mid 1)^p\big) \mid$$

$$\big([action](0 \mid 1)^p(0 \mid 1)^{n-m-1}1(0 \mid 1)^{m-1}0(0 \mid 1)^p[action](0 \mid 1)^p(0 \mid 1)^{n-m-1}1(0 \mid 1)^{m-1}1(0 \mid 1)^p\big)\bigg)^*$$

$$[even] \mid [odd]\big)^*$$

The idea is that each substring $[i]$ marking an even position in the tape has to be followed by it's successor $[i]$, by forcing the word to be a concatenation of consecutive constructs $action\,statea[i]stateb\,action\,statea[j]stateb$ in which (1) $[i]$ ends in 0 and $[j]$ ends in 1, and (2) the $t$-th bit of $[i]$ and $[j]$ are equal, for every $1 \leq t \leq n-1$ (where the last bit of $[i]$ is the $n$-th one). In the same fashion, $E_{order}^{odd}$ ensures that each string representing an odd position is to be followed by it's successor, and that the last of these substrings has to be $[2^n - 1]$, or $1^n$. We omit the description since it follows the same lines of $E_{order}^{even}$.

Next, $E_i$ ensures that the first description corresponds to the initial configuration of $\mathcal{M}$: it contains the expression

$$[even][nothing][q, a_0][0][q, a_0][nothing][a_1][1][a_1][nothing][a_{k-1}][k-1][a_{k-1}]$$
$$\bigl([nothing][B](0 \mid 1)^n[B]\bigr) * [even]\Sigma_{\mathcal{M}}^*$$

Here we abuse the notation, and put $[a]$ instead of the string coding the content $a \in \Sigma$, and $[q, a]$ instead of the string coding the content $a \in \Sigma_{\mathcal{M}}$, and stating that the head is in the corresponding position, in a state $q \in Q_{\mathcal{M}}$.

Furthermore, the expressions in $E_f$ ensure that the last description ends in a final state: it contains only the expression

$$\bigcup_{q \in Q, a \in \Gamma} \Sigma^*[even] \mid [odd]$$
$$\bigl([action](0 \mid 1)^p (0 \mid 1)^n (0 \mid 1)^p\bigr)^* [head](0 \mid 1)^p (0 \mid 1)^n [q, a]\bigl([action](0 \mid 1)^p (0 \mid 1)^n (0 \mid 1)^p\bigr)^*$$
$$[even] \mid [odd]$$

Next, we ensure that the state is carried along the descriptions: the state after the action in the $j$-th description has to coincide with the state carried along the $j + 1$-th description. This is accomplished with parameterized expressions $E_{state}^{even}$ and $E_{state}^{odd}$, where $E_{state}^{even}$ requires that, for all $i$, $0 \leq i \leq 2^n - 1$, the state after an even computation in the $i$-th cell of the tape corresponds exactly to the state before the next computation.

$$E_{state}^{even} = \Biggl( \bigcup_{state \in \Gamma \cup \Gamma \times Q} \bigl([even]([action](0 \mid 1)^p (0 \mid 1)^n (0 \mid 1)^p\bigr)^*$$
$$([action](0 \mid 1)^p x_1 \cdots x_n[state]([action](0 \mid 1)^p (0 \mid 1)^n (0 \mid 1)^p)^*[even]$$
$$[odd]([action](0 \mid 1)^p (0 \mid 1)^n (0 \mid 1)^p)^*([action][state]x_1 \cdots x_n (0 \mid 1)^p$$
$$([action](0 \mid 1)^p (0 \mid 1)^n (0 \mid 1)^p)^*[odd]) \Biggr)^*,$$

Expression $E_{state}^{odd}$ is defined accordingly, simply by interchanging strings $[even]$ and $[odd]$, but carefully allowing for the possibility that a word representing a computation might end in an odd configuration (that is, an odd configuration may be followed by an even configuration with the aforementioned properties, or may be the last configuration of the computation).

Finally, we describe the set $E_{action}$. Intuitively, the expressions in this set ensure that in each configuration a step is taken that is valid w.r.t. the transitions in $\delta_{\mathcal{M}}$. Roughly speaking, it forces each configuration to be of form $L_\delta$, for some transition $\delta \in \delta_{\mathcal{M}}$. The formal description is as follows. For each transition in $\delta_{\mathcal{M}}$ of form $\delta(q, a) = (q', a', \{\rightarrow, \leftarrow\})$, let $L_{delta}$ be the language

$$\bigcup_{b \in \Gamma} \Big( [even] \mid [odd] \big( [nothing](0 \mid 1)^p (0 \mid 1)^n (0 \mid 1)^p \big)^*$$

$$[read][q,a](0 \mid 1)^n[a'][head][b](0 \mid 1)^n[q',b]\big([nothing](0 \mid 1)^p(0 \mid 1)^n(0 \mid 1)^p\big)^*[even] \mid [odd]\Big),$$

if $\delta$ advances to the right, or the language

$$\bigcup_{b \in \Gamma} \Big( [even] \mid [odd] \big( [nothing](0 \mid 1)^p (0 \mid 1)^n (0 \mid 1)^p \big)^*$$

$$[head][b](0 \mid 1)^n[q',b][read][q,a](0 \mid 1)^n[a']\big([nothing](0 \mid 1)^p(0 \mid 1)^n(0 \mid 1)^p\big)^*[even] \mid [odd]\Big),$$

if $\delta$ moves to the left. Then, $E_{action}$ includes the expression

$$[even]\big([nothing](0 \mid 1)^p(0 \mid 1)^n(0 \mid 1)^p\big)^*[even]\big(\bigcup_{\delta \in \delta_{\mathcal{M}}} L_\delta\big)^*.$$

Moreover, it also includes the expression

$$([even] \mid [odd])\Big(\big(\bigcup_{state \in \Gamma \cup \Gamma \times Q} [nothing][state](0 \mid 1)^n[state]\big) \mid$$

$$\big([read](0 \mid 1)^p(0 \mid 1)^n(0 \mid 1)^p\big) \mid$$

$$\big([head](0 \mid 1)^p(0 \mid 1)^n(0 \mid 1)^p\big)\Big)^*[even] \mid [odd])^*,$$

which ensures that the content of a cell does not change if no action is perceived on it.

It follows immediately from the construction that $\bigcap_{e \in E} \mathcal{L}_\square(e)$ is empty if and only if $\mathcal{M}$ accepts on input $\bar{a}$.

(Part 2) Let $\mathcal{A}$ be an NFA over $\Sigma \cup \mathcal{W}$, where $\mathcal{W}$ is a set of variables. Then, for every two valuations $\nu_1$ and $\nu_2$, we have $\mathcal{L}(\nu_1(\mathcal{A})) \neq \emptyset$ iff $\mathcal{L}(\nu_2(\mathcal{A})) \neq \emptyset$. Indeed, a path from an initial to a final state in one automaton guarantees the existence of such a path in the other one. Hence to check $\mathcal{L}_\diamond(\mathcal{A}) \neq \emptyset$ one can take an arbitrary valuation $\nu$ and check whether $\mathcal{L}(\nu(\mathcal{A})) \neq \emptyset$; this of course is reachability and can be done in NLOGSPACE. Hardness is already known for automata that do not use variables.

### Proof of Lemma 2 :

Consider first the case when $\Sigma$ has at least two symbols. Then, by using the construction of Lemma 3 it is easy to see that the resulting expression $e'$ is of size $O(|\Sigma| \cdot k + k \cdot (|\Sigma| + n + k))$. We have assumed than $k \leq n$, and it is reasonable to assume that $|\Sigma| \leq k \leq n$, thus the resulting expression is of size $O(k \cdot n)$.

If $\Sigma$ contains a single symbol, then for $1 \leq i \leq k$ we have that $\mathcal{L}_\square(e_i) = L(e'_i)$, where $e'_i$ is the expression resulting from replacing all parameters in $e_i$ with the symbol from $\Sigma$. Afterwards, $\Sigma$ can be augmented with an extra symbol and the construction of Lemma 3 can be used just as we have previously explained, except this time on input $e'_1, \ldots, e'_k$.

## Proof of Corollary 1 :

We begin the proof with some observations. Given a Turing machine $\mathcal{M}$ that decides on input $\bar{a}$ using exponential space, the reduction of Theorem 1 essentially constructs a set of parameterized regular expressions $E_{\mathcal{M},\bar{a}}$ such that every word $w \in \bigcap_{e \in E_{\mathcal{M},\bar{a}}} \mathcal{L}_\square(e)$ represents a run (or, more precisely, a sequence of configurations) of $\mathcal{M}$ on input $a$.

Moreover, the same proof also shows how to construct a regular expression $e_{\mathcal{M},\bar{a}}$ such that $\bigcap_{e \in E_{\mathcal{M},\bar{a}}} \mathcal{L}_\square(e)$ is empty if and only if $\mathcal{L}_\square(e_{\mathcal{M},\bar{a}})$ is empty. Notice as well that from the proof of Lemma 3 it is easy to see that the size of every word $w \in \mathcal{L}_\square(e_{\mathcal{M},\bar{a}})$ is at least the size of the smallest word $w' \in \bigcap_{e \in E_{\mathcal{M},\bar{a}}} \mathcal{L}_\square(e)$.

Next, we define the sequence $\{e_n\}_{n \in \mathbb{N}}$ of parameterized regular expressions as stated in the corollary. Clearly, for each $n \in \mathbb{N}$, it is possible to construct a deterministic Turing machine $\mathcal{M}_n$ over alphabet $\Sigma = \{0,1\}$ that on input $1^n$ works for exactly $2^{2^n}$ steps, using at most $2^n$ cells.

Thus, for each $n$, we define $e_n$ as the expression such that $\mathcal{L}_\square(e_n)$ is empty if and only if $\bigcap_{e \in E_{\mathcal{M}_n, 1^n}} \mathcal{L}_\square(e)$ is empty, constructed as in the proof of Lemma 3.

According to the reduction in the proof of Theorem 1, $\bigcap_{e \in E_{\mathcal{M}_n, 1^n}} \mathcal{L}_\square(e)$ contains a single word, of length greater than $2^{2^n}$, representing the single run of $\mathcal{M}_n$ on input $1^n$. Then, by our above observations, we conclude that all words in $\mathcal{L}_\square(e_n)$ are of size at least $2^{2^n}$.

## Proof of Proposition 1 :

(Part 1) The EXPSPACE reduction on Theorem 1 can be modified so that all the expressions in $E$ are simple. In order to do this, we shall modify the coding scheme of the previous reduction. Basically, the new reduction should include one preceding bit in each of the slots corresponding to *action*, *stateb* (content of the cell before the action), *position* and *statea* (content of the cell after the action). This extra bit is set to 0 if these are part of an *even* configuration (that is, a configuration between two strings [*even*]), or 1 if they are part of an *odd* configuration.

Under this modified coding, all that is left to do is to modify the set $E$ according to the new coding. The only substantial change is in the sets $E_{state}^{even}$ and $E_{state}^{odd}$, the rest being straightforward. Thus, we only state these expressions. Define $E_{state}^{even}$ as follows:

$$\left(\bigcup_{state \in \Gamma \cup \Gamma \times Q} ([even]((0 \mid 1)^4 (0 \mid 1)^{p+1} (0 \mid 1)^{n+1} (0 \mid 1)^{p+1})^* ((0 \mid 1)^4 (0 \mid 1)^{p+1}\right.$$
$$\left(x_1 \cdots x_n \big(0[state](0(0 \mid 1)^3 0(0 \mid 1)^p (0 \mid 1)^n 0(0 \mid 1)^p)^* [even]\right.$$
$$[odd](1(0 \mid 1)^3 1(0 \mid 1)^p (0 \mid 1)^n 1(0 \mid 1)^p)^* 1(0 \mid 1)^3 1[state]) \mid$$
$$\left.\left.\big(1(0 \mid 1)^p 1(0 \mid 1)^3 1(0 \mid 1)^p (0 \mid 1)^n 1(0 \mid 1)^p)^* [odd]\big)\right)^*\right)^*.$$

And, $E_{state}^{odd}$ is defined accordingly, by mutually interchanging all appearances of [*even*] and [*odd*].

Notice that, from the form of the rest expressions in $E$, the only words $w \in \mathcal{L}_\square(E_{state}^{even})$ that belongs to $\bigcap_{e \in E} \mathcal{L}_\square(e)$ are those in which after every string of form $\nu(x_1) \cdots \nu(x_n)$ for some valuation

$\nu : \Sigma \to \{x_1, \ldots, x_n\}$ we follow with a string in

$$0[state](0(0 \mid 1)^3 0(0 \mid 1)^p (0 \mid 1)^n 0(0 \mid 1)^p)^*[even]$$
$$[odd](1(0 \mid 1)^3 1(0 \mid 1)^p (0 \mid 1)^n 1(0 \mid 1)^p)^*$$
$$1(0 \mid 1)^3 1[state]\nu(x_1) \cdots \nu(x_n) 1(0 \mid 1)^p$$
$$1(0 \mid 1)^3 1(0 \mid 1)^p (0 \mid 1)^n 1(0 \mid 1)^p)^*[odd] \quad (1)$$

Thus, under this modified coding, expressions $E_{state}^{even}$ and $E_{state}^{odd}$ have the same intended behavior as in the reduction of Theorem 1. We omit the rest of the reduction since it goes along the same lines as the previous version.

(Part 2): It is easy to see that, if $e$ does not use Kleene star, then all the words $w \in \mathcal{L}_\square(e)$ are of size polynomial with respect to the size of $e$. This immediately gives a $\Sigma_2^P$ algorithm for the emptiness problem: Given a a parameterized regular expression $e$ not using Kleene star, guess a word $w$ and a valuation $\nu$ for $e$, and check that $w \notin \mathcal{L}_\square(e)$ (which, from Theorem 2, can be performed using an NP oracle). The $\Sigma_2^P$ hardness is established via a reduction from the compliment of the $\forall\exists$ 3-SAT satisfiability problem. This problem is defined as follows: formula $\varphi$ is the conjunction of clauses $\{C_1, \ldots, C_p\}$, each of which has 3 variables from the disjoint union of $\{x_1, \ldots, x_m\}$ and $\{y_1, \ldots, y_t\}$. The problem asks whether there exists an assignment $\sigma_{\bar{x}}$ for $\{x_1, \ldots, x_m\}$ such that for every assignment $\sigma_{\bar{y}}$ for $\{y_1, \ldots, y_t\}$ it is the case that $\varphi$ is not satisfiable.

Let $\varphi := \forall x_1 \cdots \forall x_m \exists y_1 \ldots \exists y_t \, C_1 \wedge \cdots \wedge C_p$ be an instance of $\forall\exists$ 3-SAT. From $\varphi$ we construct in polynomial time a parameterized regular expression $e$ over alphabet $\Sigma = \{0, 1\}$ such that there exists an assignment $\sigma_{\bar{x}}$ for $\{x_1, \ldots, x_m\}$ such that for every assignment $\sigma_{\bar{y}}$ for $\{y_1, \ldots, y_t\}$ it is the case that $\varphi$ is not satisfiable if and only if $\mathcal{L}_\square(e)$ is not empty.

Let each $C_j$ ($1 \leq j \leq p$) be of form $(\ell_j^1 \vee \ell_j^2 \vee \ell_j^3)$, where each literal $\ell_j^i$, for $1 \leq j \leq p$ and $1 \leq i \leq 3$, is either a variable in $\{x_1, \ldots, x_m\}$ or $\{y_1, \ldots, y_t\}$, or its negation. We associate with each propositional variable $x_k$, $1 \leq k \leq m$, a fresh variable $X_k$ (representing the positive literal) and a fresh variable $\hat{X}_k$ (representing the negation of such literal). Also, with each propositional variable $y_k$, $1 \leq k \leq t$, we associate fresh variables $Y_k$ and $\hat{Y}_k$. Then let $\mathcal{W} = \{X_1, \ldots, X_m, \hat{X}_1, \ldots, \hat{X}_m\} \cup \{Y_1, \ldots, Y_t, \hat{Y}_1, \ldots, \hat{Y}_t\} \cup \{Z\}$, where $Z$ is a fresh variable as well.

Next we define expression $e$ over $\Sigma = \{0, 1\}$ and $\mathcal{W}$. We find it useful to split the definition in two parts. First, define
$$e = (Z \cdot 0 \cdot e_1) \mid (1Ze_2),$$
with
$$e_1 = 1100(0 \mid 1)^m 000$$
and
$$e_2 = e_{2,2,1} \mid \cdots \mid e_{2,2,m} \mid e_{2,2,1} \mid \cdots \mid e_{2,2,m} \mid e_{2,3,1} \mid \cdots \mid e_{2,3,t} \, e_{2,3} \mid e_{2,4},$$
where

- for each $1 \leq k \leq m$, define $e_{2,1,k} = 1100(0 \mid 1)^{k-1} X_k (0 \mid 1)^{m-k} 000$

- for each $1 \leq k \leq m$, define $e_{2,2,k} = \big(X_k \hat{X}_k 00 (0 \mid 1)^m 000\big) \mid \big(11 X_k \hat{X}_k (0 \mid 1)^m 000\big)$;

- for each $1 \leq k \leq t$, $e_{2,3,k} = \big(Y_k \hat{Y}_k 00 (0 \mid 1)^m 000\big) \mid \big(11 Y_k \hat{Y}_k (0 \mid 1)^m 000\big)$;

- Let $h$ be a function that maps each literal $\ell_j^i$ to the variable $X_k$, if $\ell_j^i$ corresponds to $x_k$ or to $\hat{X}_k$, if $\ell_j^i$ corresponds to $\neg x_k$ ($1 \leq j \leq p$, $1 \leq i \leq 3$ and $1 \leq k \leq m$); or to $Y_k$, if $\ell_j^i$ corresponds to $y_k$ or to $\hat{Y}_k$, if $\ell_j^i$ corresponds to $\neg y_k$. Then define $e_{2,4} = 1100(0 \mid 1)^m \big( h(\ell_1^1) \cdot h(\ell_1^2) \cdot h(\ell_1^3) \mid \cdots \mid h(\ell_p^1) \cdot h(\ell_p^2) \cdot h(\ell_p^3) \big)$.

We now prove that $\mathcal{L}_\square(e) \neq \emptyset$ if and only if there exists an assignment $\sigma_{\bar{x}}$ for $\{x_1, \ldots, x_m\}$ such that for every assignment $\sigma_{\bar{y}}$ for $\{y_1, \ldots, y_t\}$ it is the case that $\varphi$ is not satisfiable.

($\Leftarrow$): Assume first that there exists an assignment $\sigma_{\bar{x}}$ for $\{x_1, \ldots, x_m\}$ such that for every assignment $\sigma_{\bar{y}}$ for $\{y_1, \ldots, y_t\}$ it is the case that $\varphi$ is not satisfiable. Define $a_1, \ldots, a_m$ as follows: for each $1 \leq k \leq m$, $a_k = 0$ if and only iff $\sigma_{\bar{x}}$ assigns the value 1 to the variable $x_k$. Let then $w = 101100 a_1 \cdots a_m 000$ be a word in $\Sigma^*$. We claim that $w \in \mathcal{L}_\square(e)$. To prove this claim, let $\nu : \mathcal{W} \to \Sigma$ be a valuation. We show that $w \in \mathcal{L}(\nu(e))$. The proof is done via a case by case analysis:

- Assume first that $\nu(Z) = 1$. Then, since $1100 a_1 \cdots a_m 000$ is clearly denoted by $e_1$, we have that $w \in \mathcal{L}(\nu(e))$.

- Next, assume that $\nu(Z) = 0$, and for some $1 \leq k \leq m$ it is the case that $\nu(X_k) = \nu(\hat{X}_k)$. Then, it is easy to see that $1100 a_1 \cdots a_m 000$ is denoted by expression

$$\nu(e_{2,2,k}) = \big(\nu(X_k)\nu(\hat{X}_k)00(0 \mid 1)^m 000\big) \mid \big(11\nu(X_k)\nu(\hat{X}_k)(0 \mid 1)^m 000\big).$$

And thus we have that $w \in \mathcal{L}(\nu(e))$.

- Assume now that $\nu(Z) = 0$, and for some $1 \leq k \leq t$ it is the case that $\nu(Y_k) = \nu(\hat{Y}_k)$. Then, it is easy to see that $1100 a_1 a_1 \cdots a_m a_m 000$ is denoted by expression

$$\nu(e_{2,3,k}) = \big(\nu(Y_k)\nu(\hat{Y}_k)00(0 \mid 1)^m 000\big) \mid \big(11\nu(Y_k)\nu(\hat{Y}_k)(0 \mid 1)^m 000\big).$$

And thus we have that $w \in \mathcal{L}(\nu(e))$.

The remaining valuations are such that $\nu(Z) = 0$, $\nu(X_k) \neq \nu(\hat{X}_k)$ for each $1 \leq k \leq m$, and for each $1 \leq k \leq t$ we have that $\nu(Y_k) \neq \nu(\hat{Y}_k)$. Let us continue.

- Assume now that for some $1 \leq k \leq m$ it is the case that $\nu(X_k) = 1$ but $\sigma_{\bar{x}}(x_k) = 0$. Then, $a_k = 1$, and thus it is easy to see that the word $1100 a_1 \cdots a_m 000$ is denoted by $\nu(e_{2,1,k})$, that corresponds to the expression $1100(0 \mid 1)^{k-1} \nu(X_k)(0 \mid 1)^{m-k} 000$ which entails that $w$ is denoted by $\nu(e)$.

- Assume now that that $\nu(Z) = 0$, $\nu(X_k) \neq \nu(\hat{X}_k)$ for each $1 \leq k \leq m$, for each $1 \leq k \leq t$ we have that $\nu(Y_k) \neq \nu(\hat{Y}_k)$, and for some $1 \leq k \leq m$ it is the case that $\nu(X_k) = 0$ but $\sigma_{\bar{x}}(x_k) = 1$. Then, $a_k = 0$, and thus it is easy to see that the word $1100 a_1 \cdots a_m 000$ is denoted by $\nu(e_{2,1,k})$, that corresponds to the expression $1100(0 \mid 1)^{k-1} \nu(X_k)(0 \mid 1)^{m-k} 000$ which entails that $w$ is denoted by $\nu(e)$.

- Finally, assume that $\nu(Z) = 0$, $\nu(X_k) \neq \nu(\hat{X}_k)$ for each $1 \leq k \leq m$, for each $1 \leq k \leq t$ we have that $\nu(Y_k) \neq \nu(\hat{Y}_k)$, and for each $1 \leq k \leq m$ it is the case that $\nu(X_k) = \sigma_{\bar{x}}(x_k)$. Define the following valuation $\sigma_{\bar{y}}$ for the variables in $\{y_1, \ldots, y_t\}$: $\sigma_{\bar{y}}(y_k) = \nu(Y_k)$, for each $1 \leq k \leq t$. From our initial assumption, there exists at least a clause $C_j$, $1 \leq j \leq p$, that is falsified under the assignment $\sigma_{\bar{x}}, \sigma_{\bar{y}}$. From the fact that $\nu(X_k) \neq \nu(\hat{X}_k)$ for each $1 \leq k \leq m$, for each $1 \leq k \leq t$ we have that $\nu(Y_k) \neq \nu(\hat{Y}_k)$, and for each $1 \leq k \leq m$ it is the case that $\nu(X_k) = \sigma_{\bar{x}}(x_k)$, we conclude that $\nu(h(\ell_1^1)) \cdot \nu(h(\ell_1^2)) \cdot \nu(h(\ell_1^3))$ corresponds to the string 000, which proves that $1100a_1 \cdots a_m 000$ belongs to $\mathcal{L}(\nu(e_{2,4}))$, and thus $w$ is denoted by $\nu(e)$.

($\Rightarrow$): Assume now that there exists a word $w$ in $\mathcal{L}_\square(e)$. It is straightforward to show that $w$ must begin with the prefix 10: if $w$ begins with either 00 or 01 then it cannot be denoted by $\nu_1(e)$, where $\nu_1$ assigns a letter 1 to all variables in $\mathcal{W}$. Moreover, if it begins with 11 then it cannot be accepted by the valuation that assigns the letter 0 to all variables. Let then $w = 10v$. Furthermore, let $\nu$ be an arbitrary valuation such that $\nu(Z) = 1$. since $w$ belongs to $\mathcal{L}(\nu(e))$, then $v$ must be denoted by $e_1$, and so it must be that $v$ is of form $1100(0 \mid 1)^m 000$. Let $\sigma_{\bar{x}}$ be a valuation for $\{x_1, \ldots, x_m\}$ defined as follows: $\sigma_{\bar{x}}(x_k) = 1$ if $v$ is of form $1100(0 \mid 1)^{k-1}0(0 \mid 1)^{m-k}000$, and $\sigma_{\bar{x}}(x_k) = 0$ if $v$ is of form $1100(0 \mid 1)^{k-1}1(0 \mid 1)^{m-k}000$. Next, we show that for each valuation $\sigma_{\bar{y}}$ for $\{y_1, \ldots, y_t\}$ it is the case that $\varphi$ is not satisfied with valuation $\sigma_{\bar{x}}, \sigma_{\bar{y}}$. Assume for the sake of contradiction that there is a valuation $\sigma_{\bar{y}}$ for $\{y_1, \ldots, y_t\}$ such that $\sigma_{\bar{x}}, \sigma_{\bar{y}}$ satisfies $\varphi$. Define the following valuation $\nu : \mathcal{W} \to \Sigma$:

- $\nu(Z) = 0$

- $\nu(X_k) = \sigma_{\bar{x}}(x_k)$, for each $1 \leq k \leq m$,

- $\nu(Y_k) = \sigma_{\bar{y}}(y_k)$, for each $1 \leq k \leq t$,

- $\nu(\hat{X}_k) = 1$ if and only if $\nu(X_k) = 0$,

- $\nu(\hat{Y}_k) = 1$ if and only if $\nu(Y_k) = 0$

Let us now show that $10v \notin \mathcal{L}(\nu(e))$, contradicting our initial assumption that $\mathcal{L}_\square(e) = \emptyset$. Since $\nu(Z) = 0$, we have then that $v$ must be denoted by $e_2$. Since $\nu(X_k) \neq \nu(\hat{X}_k)$ for each $1 \leq k \leq m$ and for each $1 \leq k \leq t$ we have that $\nu(Y_k) \neq \nu(\hat{Y}_k)$, it is easy to see that word $v$ cannot be in $\mathcal{L}(\nu(e_{2,2,k}))$ for $1 \leq k \leq m$ or in $\mathcal{L}(\nu(e_{2,3,k}))$ for $1 \leq k \leq t$. Moreover, since $\sigma_{\bar{x}}(x_k) = 1 = \nu(X_k)$ if $v$ is of form $1100(0 \mid 1)^{k-1}0(0 \mid 1)^{m-k}000$, and $\sigma_{\bar{x}}(x_k) = 1 = \nu(Y_k)$ if $v$ is of form $1100(0 \mid 1)^{k-1}0(0 \mid 1)^{m-k}000$, we have that $v$ cannot be in $\mathcal{L}(\nu(e_{2,1,k}))$ for $1 \leq k \leq m$. The only remaining possibility is that the word $v$ belongs to $\nu(e_4)$. This implies that for some $1 \leq j \leq p$, it is the case that $\nu(h(\ell_1^1)) \cdot \nu(h(\ell_1^2)) \cdot \nu(h(\ell_1^3))$ corresponds to the string 000 But we know from the definition of $\nu$ that that cannot be true, as we have assumed that $\sigma_{\bar{x}}, \sigma_{\bar{y}}$ satisfies $\varphi$. We conclude that $10v \notin \mathcal{L}(\nu(e))$, which was to be shown.

### Proof of Theorem 2:

The first part follows from the first part of Theorem 3. We prove the second part here. We use a reduction from POSITIVE 1-3 3-SAT, which is the following NP-hard decision problem: Given a conjunction $\varphi$ of clauses, with exactly three literals each, and in which no negated variable occurs, is there a truth assignment to the variables so that each clause has exactly one true variable?

The reduction is as follows. Let $\varphi = C_1 \wedge \cdots \wedge C_m$ be a formula in CNF, where each $C_i$ ($1 \leq i \leq m$) is a clause consisting of exactly tree positive literals. Let $\{p_1, \ldots, p_n\}$ be the variables that appear in $\varphi$. With each propositional variable $p_i$ ($1 \leq i \leq n$) we associate a different variable $x_i \in \mathcal{V}$. We show next how to construct, in polynomial time from $\varphi$, a parameterized regular expression $e$ over alphabet $\Sigma = \{a, 0, 1\}$ and a word $w$ over the same alphabet, such that there is an assignment to the variables of $\varphi$ for which each clause has exactly one true variable if and only if $w \in \mathcal{L}_\diamond(e)$.

The parameterized regular expression $e$ is defined as $ae_1ae_2a\cdots ae_ma$, where the regular expression $e_i$, for $1 \leq i \leq m$, is defined as follows: Assume that $C_i = (p_j \vee p_k \vee p_\ell)$, where $1 \leq j, k, \ell \leq n$. Then $e_i$ is defined as $(x_jx_kx_\ell \,|\, x_jx_\ell x_k \,|\, x_kx_jx_\ell \,|\, x_kx_\ell x_j \,|\, x_\ell x_jx_k \,|\, x_\ell x_kx_j)$. That is, $e_i$ is just the union of all the possible forms in which the variables in $\mathcal{V}$ that correspond to the propositional variables that appear in $C_i$ can be ordered. Further, the word $w$ is defined as $(a100)^ma$. Clearly, $e$ and $w$ can be constructed in polynomial time from $\varphi$. Next we show that there is an assignment for variables $\{p_1, \ldots, p_n\}$ for which each clause has exactly one true variable if and only if $w \in \mathcal{L}_\diamond(e)$.

Assume first that $w \in \mathcal{L}_\diamond(e)$. Then there exists a valuation $\nu : \{x_1, \ldots, x_n\} \to \Sigma$ such that $w \in \mathcal{L}(\nu(e))$. Thus, it must be the case that the word $a100$ belongs to $\nu(ae_i)$, for each $1 \leq i \leq m$. But this implies that if $C_i = (x_j \vee x_k \vee x_\ell)$, then $\nu$ assigns value 1 to exactly one of the variables in the set $\{x_j, x_k, x_\ell\}$ and it assigns value 0 to the other two variables. Let us define now a propositional assignment $\sigma : \{p_1, \ldots, p_n\} \to \{0, 1\}$ such that $\sigma(p_i) = \nu(x_i)$, for each $1 \leq i \leq n$. It is not hard to see then that for each clause $C_j$, $1 \leq j \leq m$, $\sigma$ assigns value 1 to exactly one of its propositional variables.

Assume, on the other hand, that there is a propositional assignment $\sigma : \{p_1, \ldots, p_n\} \to \{0, 1\}$ that assigns value 1 to exactly one variable in each clause $C_i$, $1 \leq i \leq m$. Let us define $\nu$ as a valuation from $\{x_1, \ldots, x_n\}$ into $\{0, 1\}$ such that $\nu(x_i) = 1$ if and only if $\sigma(p_i) = 1$. Clearly then $100 \in \mathcal{L}(\nu(e_i))$, for each $1 \leq i \leq m$. Thus, $(a100)^ma \in \mathcal{L}(\nu(e))$. We conclude that $w \in \mathcal{L}_\diamond(e)$.

## Proof of Proposition 2:

For the sake of readability, in this proof we use $\cup$ – instead of $|$ – for representing the operation of union between regular expressions.

We first consider the $\diamond$-semantics. Notice that the reduction used in the proof of Theorem 2, to show NP-hardness of MEMBERSHIP$_\diamond$, constructs a regular expression that is of star-height 0. This shows NP-hardness of MEMBERSHIP$_\diamond$ for expressions of star height 0. We prove NP-hardness of MEMBERSHIP$_\diamond$ for simple expressions here.

We use a reduction from 3-SAT. Let $\varphi = \bigwedge_{1 \leq i \leq n}(\ell_i^1 \vee \ell_i^2 \vee \ell_i^3)$ be a propositional formula in 3-CNF over variables $\{p_1, \ldots, p_m\}$. That is, each literal $\ell_i^j$, for $1 \leq i \leq n$ and $1 \leq j \leq 3$, is either $p_k$ or $\neg p_k$, for $1 \leq k \leq m$. Next, we show how to construct in polynomial time from $\varphi$, a simple regular expression $e$ over alphabet $\Sigma = \{a, b, c, d, 0, 1\}$ and a word $w$ over the same alphabet such that $\varphi$ is satisfiable if and only if $w \in \mathcal{L}_\diamond(e)$.

The regular expression $e$ is defined as $f^*$, where $f := a(f_1 \cup g_1 \cup \cdots \cup f_m \cup g_m)b$, and the regular expressions $f_i$ and $g_i$ are defined as follows.

Intuitively, $f_i$ (resp. $g_i$) codifies $p_i$ (resp. $\neg p_i$) and the clauses in which $p_i$ (resp. $\neg p_i$) appears. Formally, we define $f_i$ ($1 \leq i \leq m$) as $((c^i \cup \bigcup_{\{1 \leq j \leq n | p_i = \ell_j^1 \text{ or } p_i = \ell_j^2 \text{ or } p_i = \ell_j^3\}} d^j) \cdot x_i)$, where $x_i$ is a fresh variable in $\mathcal{V}$. In the same way we define $g_i$ as $((c^i \cup \bigcup_{\{1 \leq j \leq n | \neg p_i = \ell_j^1 \text{ or } \neg p_i = \ell_j^2 \text{ or } \neg p_i = \ell_j^3\}} d^j) \cdot \bar{x}_i)$, where $\bar{x}_i$ is a fresh variable in $\mathcal{V}$. The variable $x_i$ (resp. $\bar{x}_i$) is said to be *associated* with $p_i$ (resp. $\neg p_i$) in $e$.

Clearly, $e$ is a simple regular expression and can be constructed in polynomial time from $\varphi$.

The word $w$ is defined as:

$$ac1b\,ac0b\,acc1b\,acc0b\cdots ac^m1b\,ac^m0b\,ad1b\,add1b\cdots ad^n1b.$$

Clearly, $w$ can be constructed in polynomial time from $\varphi$. Next we show that $\varphi$ is satisfiable if and only if $w \in \mathcal{L}_\diamond(e)$.

Assume first that $w \in \mathcal{L}_\diamond(e)$. That is, there is a valuation $\nu$ for the variables $\{x_1, \bar{x}_1, \ldots, x_m, \bar{x}_m\}$ over $\Sigma$ such that $w \in \mathcal{L}(\nu(e))$. But then, given the form of $w$, it is clear that $ac^i 1 b$ and $ac^i 0 b$ belong to $\mathcal{L}(\nu(f))$, for each $1 \leq i \leq m$. Notice that the only way for this happen is that both $\nu(x_i)$ and $\nu(\bar{x}_i)$ take its value in the set $\{0, 1\}$, and, further, $\nu(x_i) \neq \nu(\bar{x}_i)$. For the same reasons, $ad^j 1 b \in \mathcal{L}(\nu(f))$, for each $1 \leq j \leq n$. But the only way for this to happen is that for each $1 \leq j \leq n$ it is the case that either the variable associated with $\ell_j^1$ or with $\ell_j^2$ or with $\ell_j^3$ in $e$ is assigned value 1 by $\nu$. Thus, the propositional assignment $\sigma : \{p_1, \ldots, p_m\} \to \{0, 1\}$, defined as $\sigma(p_i) = 1$ if and only if $\nu(x_i) = 1$, is well-defined and satisfies $\varphi$.

Assume, on the other hand, that there is a satisfying propositional assignment $\sigma : \{p_1, \ldots, p_m\} \to \{0, 1\}$ for $\varphi$. Consider the following valuation $\nu : \{x_1, \bar{x}_1, \ldots, x_m, \bar{x}_m\}$ for $e$: $\nu(x_i) = \sigma(p_i)$ and $\nu(\bar{x}_i) = 1 - \sigma(x_i)$. Using essentially the same techniques than in the previous paragraph it is possible to show that $w \in \mathcal{L}(\nu(e))$, and, therefore, that $w \in \mathcal{L}_\diamond(e)$.

Next we show that MEMBERSHIP$_\diamond$ can be solved in time $O(mn \cdot \log^2 n)$ for simple expressions of star-height 0. Given a regular expression $e \in \text{REG}(\Sigma, \mathcal{V})$ that is simple and of star-height 0, one can construct in time $O(n \cdot \log^2 n)$ [13] an $\varepsilon$-free NFA $\mathcal{A}$ over $\Sigma \cup \mathcal{V}$ that accepts precisely $\mathcal{L}(e)$, and satisfies the following two properties: (1) Its underlying directed graph is simple and acyclic (this is because $e$ does not mention the Kleene star), and (2) for each $x \in \mathcal{V}$ that is mentioned in $e$ there is at most one pair $(q, q')$ of states of $\mathcal{A}$ such that $\mathcal{A}$ contains a transition from $q$ to $q'$ labeled $x$ (this is because $e$ is simple). From Lemma 1, checking whether $w \in \mathcal{L}_\diamond(e)$, for a given word $w \in \Sigma^*$, is equivalent to checking whether $w \in \mathcal{L}(\nu(\mathcal{A}))$, for some valuation $\nu$ for $\mathcal{A}$. We show how the latter can be done in polynomial time.

First, construct in time $O(m)$ a DFA $\mathcal{B}$ over $\Sigma$ such that $\mathcal{L}(\mathcal{B}) = \{w\}$. We assume without loss of generality that the set $Q$ of states of $\mathcal{A}$ is disjoint from the set $P$ of states of $\mathcal{B}$. Next we construct, the following NFA $\mathcal{A}'$ over alphabet $\Sigma \cup (\mathcal{V} \times \Sigma)$ as follows: The set of states of $\mathcal{A}'$ is $Q \times P$. The initial state of $\mathcal{A}'$ is the pair $(q_0, p_0)$, where $q_0$ is the initial state of $\mathcal{A}$ and $p_0$ is the initial state of $\mathcal{B}$. The final states of $\mathcal{A}'$ are precisely the pairs $(q, p) \in Q \times P$ such that $q$ is a final state of $\mathcal{A}$ and $p$ is a final state of $\mathcal{B}$. Finally, there is a transition in $\mathcal{A}'$ from state $(q, p)$ to state $(q', p')$ labeled $a \in \Sigma$ if and only there is a transition in $\mathcal{A}$ from $q$ to $q'$ labeled $a$ and there is a transition in $\mathcal{B}$ from $p$ to $p'$ labeled $a$. There is a transition in $\mathcal{A}'$ from state $(q, p)$ to state $(q', p')$ labeled $(x, a) \in \mathcal{V} \times \Sigma$ if and only there is a transition in $\mathcal{A}$ from $q$ to $q'$ labeled $x$ and there is a transition in $\mathcal{B}$ from $p$ to $p'$ labeled $a$. Clearly, such construction can be performed by checking all combinations of transitions of both $\mathcal{A}$ and $\mathcal{B}$, and thus it can be performed in time $O(mn \cdot \log^2 n)$. Checking whether $\mathcal{L}(\mathcal{A}') \neq \emptyset$ can easily be done in linear time w.r.t. the size of $\mathcal{A}'$, thus obtaining the $O(mn \cdot \log^2 n)$ bound. We prove next that checking this is equivalent to checking whether $w \in L(\nu(\mathcal{A}))$, for some valuation $\nu$ for $\mathcal{A}$, which finishes the proof of the proposition in terms of the $\diamond$-semantics.

Assume first that $\mathcal{L}(\mathcal{A}') \neq \emptyset$. Let $(q_0, p_0) \xrightarrow{u_1} (q_1, p_1) \xrightarrow{u_2} \cdots \xrightarrow{u_{n-1}} (q_{n-1}, p_{n-1}) \xrightarrow{u_n} (q_n, p_n)$ be an accepting run of $\mathcal{A}'$. That is, $u_1 u_2 \cdots u_n \in (\Sigma \cup (\mathcal{V} \times \Sigma))^*$ and $(q_n, p_n)$ is a final state of $\mathcal{A}'$. Since the underlying directed graph of $\mathcal{A}$ is acyclic, and each variable $x$ mentioned in $e$ appears in at most one transition of $\mathcal{A}$, it must be the case that for each $1 \leq i < j \leq n$, if $u_i = (x_i, a_i) \in \mathcal{V} \times \Sigma$ and $u_j = (x_j, a_j) \in \mathcal{V} \times \Sigma$ then $x_i \neq x_j$. This implies that we can define a mapping $\nu : \mathcal{W} \to \Sigma$, where $\mathcal{W}$ is the set of variables used in transitions of $\mathcal{A}$, such that $\nu(x) = a$, if $u_i = (x, a)$ for some $1 \leq i \leq n$, and $\nu(x)$ is an arbitrary element $a' \in \Sigma$, otherwise. It is not hard to see that $q_0 \xrightarrow{a_1} q_1 \xrightarrow{a_2} \cdots \xrightarrow{a_{n-1}} q_{n-1} \xrightarrow{a_n} q_n$ is also an accepting run of $\mathcal{L}(\nu(\mathcal{A}))$ and that $a_1 a_2 \cdots a_n = w$. The latter can be proved as follows: Let $f : \{u_1, \ldots, u_n\} \to \Sigma$ be the mapping such that $f(u_i) = u_i$, if $u_i = a \in \Sigma$, and $f(u_i) = a$, if $u_i = (x, a) \in \mathcal{V} \times \Sigma$. Then clearly $p_0 \xrightarrow{f(u_1)} p_1 \xrightarrow{f(u_2)} \cdots \xrightarrow{f(u_{n-1})} p_{n-1} \xrightarrow{f(u_n)} p_n$ is an accepting run of $\mathcal{B}$, and, therefore, $w = f(u_1) \cdots f(u_n)$. Further, let $g : \{u_1, \ldots, u_n\} \to \Sigma$ be the mapping such that $g(u_i) = u_i$, if $u_i = a \in \Sigma$, and $g(u_i) = \nu(x) = a$, if $u_i = (x, a) \in \mathcal{V} \times \Sigma$. Then clearly $f(u_i) = g(u_i)$, for each $1 \leq i \leq n$, and, further, $q_0 \xrightarrow{g(u_1)} q_1 \xrightarrow{g(u_2)} \cdots \xrightarrow{g(u_{n-1})} q_{n-1} \xrightarrow{g(u_n)} q_n$ is an accepting run of $\mathcal{L}(\nu(\mathcal{A}))$. We conclude that $w \in \mathcal{L}(\nu(\mathcal{A}))$.

Assume, on the other hand, that $w \in \mathcal{L}(\nu(\mathcal{A}))$, for some valuation $\nu$ for $\mathcal{A}$. Suppose that $w = a_1 a_2 \cdots a_n$, where each $a_i \in \Sigma$ ($1 \leq i \leq n$), and let $q_0 \xrightarrow{a_1} q_1 \xrightarrow{a_2} \cdots \xrightarrow{a_{n-1}} q_{n-1} \xrightarrow{a_n} q_n$ be an accepting run of $\mathcal{L}(\nu(\mathcal{A}))$; i.e. $q_n$ is a final state of $\mathcal{A}$. Assume that $i_1 < i_2 < \cdots < i_m$ are the only indexes in the set $\{0, 1, \ldots, n-1\}$ such that, for each $1 \leq j \leq m$, there is no transition labeled $a_{i_j+1}$ from $q_{i_j}$ to $q_{i_j+1}$ in $\mathcal{A}$. Then there must be a transition in $\mathcal{A}$ from $q_{i_j}$ to $q_{i_j+1}$ labeled $x_{i_j} \in \mathcal{V}$. Consider an arbitrary accepting run $p_0 \xrightarrow{a_1} p_1 \xrightarrow{a_2} \cdots \xrightarrow{a_{n-1}} p_{n-1} \xrightarrow{a_n} p_n$ of $\mathcal{B}$; i.e. $p_n$ is a final state of $\mathcal{B}$. Then it is clear that

$$(q_0, p_0) \xrightarrow{a_1} (q_1, p_1) \cdots (q_{i_1}, p_{i_1}) \xrightarrow{(x_{i_1}, a_{i_1+1})} (q_{i_1+1}, p_{i_1+1}) \cdots$$
$$(q_{i_m}, p_{i_m}) \xrightarrow{(x_{i_m}, a_{i_m+1})} (q_{i_m+1}, p_{i_m+1}) \cdots (q_{n-1}, p_{n-1}) \to a_n (q_n, p_n)$$

is an accepting run of $\mathcal{A}'$. Thus, $\mathcal{L}(\mathcal{A}') \neq \emptyset$.

Now we deal with the $\square$-semantics. That MEMBERSHIP$_\square$ is coNP-hard, even over the class of expressions of star-height 0, follows from Theorem 3. Next we prove that MEMBERSHIP$_\square$ is coNP-hard, even over the class of simple regular expressions.

We use a reduction from 3-SAT to the complement of MEMBERSHIP$_\square$ over the class of simple expressions. Let $\varphi = \bigwedge_{1 \leq i \leq n}(\ell_i^1 \vee \ell_i^2 \vee \ell_i^3)$ be a propositional formula in 3-CNF over variables $\{p_1, \ldots, p_m\}$. That is, each literal $\ell_i^j$, for $1 \leq i \leq n$ and $1 \leq j \leq 3$, is either $p_k$ or $\neg p_k$, for $1 \leq k \leq m$. Next, we show how to construct in polynomial time from $\varphi$, a simple regular expression $e$ over alphabet $\Sigma = \{a, b, 0, 1\}$ and a word $w$ over the same alphabet such that $\varphi$ is satisfiable if and only if $w \notin \mathcal{L}_\square(e)$.

We start by defining the word $w$ as follows:

$$w := 1111a\,11111a\,b\,1110a\,11110a\,b\,111111a\,1111111a\,b\,111110a\,1111110a\,b\,\cdots$$
$$1^{2i+1}1a\,1^{2i+2}1a\,b\,1^{2i+1}0a\,1^{2i+2}0a\,b\cdots 1^{2m+1}1a\,1^{2m+2}1a\,b\,1^{2m+1}0a\,1^{2m+2}0a\,b$$
$$1^{3(m+1)+1}0a\,1^{3(m+1)+2}0a\,1^{3(m+1)+3}0a\,b\,1^{6(m+1)+1}0a\,1^{6(m+1)+2}0a\,1^{6(m+1)+3}0a\,b\cdots$$
$$\cdots 1^{3j(m+1)+1}0a\,1^{3j(m+1)+2}0a\,1^{3j(m+1)+3}0a\,b\cdots$$
$$1^{3n(m+1)+1}0a\,1^{3n(m+1)+2}0a\,1^{3n(m+1)+3}0a\,b\,baa.$$

We denote by $w'$ the prefix of $w$ such that $w = w'aa$ and by $w''$ the prefix of $w$ such that $w = w''baa$. Clearly, $w$ can be constructed in polynomial time from $\varphi$. Next we show that $\varphi$ is satisfiable if and only if $w \notin \mathcal{L}_\square(e)$.

The regular expression $e$ is defined as $(\Sigma^* b \cup \epsilon) f (b \Sigma^* \cup \epsilon)$, where $f$ is defined as: $\big((f_1 \cup g_1 \cup \cdots f_m \cup g_m)(a \cup \epsilon)\big)^*$. Intuitively $f_i$ (resp. $g_i$) codifies $p_i$ (resp. $\neg p_i$) and the clauses in which $p_i$

(resp. $\neg p_i$) appears. Formally, we define $f_i$ $(1 \leq i \leq m)$ as

$$\left((\{w'\} \cup \{w''\} \cup 1^{2i+1} \cup \bigcup_{\{1 \leq j \leq n | p_i = \ell_j^1\}} 1^{3j(m+1)+1} \cup \bigcup_{\{1 \leq j \leq n | p_i = \ell_j^2\}} 1^{3j(m+1)+2} \cup \bigcup_{\{1 \leq j \leq n | p_i = \ell_j^3\}} 1^{3j(m+1)+3}\right) \cdot x_i a),$$

where $x_i$ is a fresh variable in $\mathcal{V}$. In the same way we define $g_i$ as

$$\left((\{w'\} \cup \{w''\} \cup 1^{2i+2} \cup \bigcup_{\{1 \leq j \leq n | \neg p_i = \ell_j^1\}} 1^{3j(m+1)+1} \cup \bigcup_{\{1 \leq j \leq n | \neg p_i = \ell_j^2\}} 1^{3j(m+1)+2} \cup \bigcup_{\{1 \leq j \leq n | \neg p_i = \ell_j^3\}} 1^{3j(m+1)+3}\right) \cdot \bar{x}_i a),$$

where $\bar{x}_i$ is a fresh variable in $\mathcal{V}$. The variable $x_i$ (resp. $\bar{x}_i$) is said to be *associated* with $p_i$ (resp. $\neg p_i$) in $e$. Clearly, $e$ is a simple regular expression and can be constructed in polynomial time from $\varphi$.

We prove first that if $w \notin \mathcal{L}_\Box(e)$ then $\varphi$ is satisfiable. Assume that $w \notin \mathcal{L}_\Box(e)$. Then there exists a valuation $\nu : \{x_1, \bar{x}_1, \ldots, x_m, \bar{x}_m\} \to \Sigma$ such that $w \notin \mathcal{L}(\nu(e))$. First of all, we prove that for each $1 \leq i \leq m$ both $\nu(x_i)$ and $\nu(\bar{x}_i)$ belong to the set $\{0, 1\}$. Assume, for the sake of contradiction, that this is not the case. Suppose first that $\nu(x_i) = a$, for some $1 \leq i \leq m$. Then it is clear that $\mathcal{L}(w'aa) \subseteq \mathcal{L}(\nu(e))$ (because $\mathcal{L}(w'\nu(x_i)a) \subseteq \mathcal{L}(\nu(e))$). But $w = w''aa$, and, therefore, $w \in \mathcal{L}(\nu(e))$, which is a contradiction. Suppose now that $\nu(x_i) = b$, for some $1 \leq i \leq m$. Then, again, it is clear that $\mathcal{L}(w''baa) \subseteq \mathcal{L}(\nu(e))$ (because $\mathcal{L}(w''\nu(x_i)aa) \subseteq \mathcal{L}(\nu(e))$). As in the previous case, $w = w''baa$, and, therefore, $w \in \mathcal{L}(\nu(e))$, which is a contradiction. The other case, when $\sigma(\bar{x}_i) \in \{a, b\}$, for some $1 \leq i \leq m$, is completely analogous.

Next we prove that for each $1 \leq i \leq m$ it is the case that $\sigma(x_i) = 1 - \sigma(\bar{x}_i)$. Assume otherwise. Then for some $1 \leq i \leq m$ it is the case that $\sigma(x_i) = \sigma(\bar{x}_i)$. Suppose first that $\sigma(x_i) = \sigma(\bar{x}_i) = 1$. Consider the unique prefix $w_1$ of $w$ that is of the form $u1^{2i+1}1a1^{2i+2}1a$, for $u \in \Sigma^*$. Then $w$ is of the form $w_1 w_2$, where $w_2 \in b\Sigma^*$. Since $w \notin \mathcal{L}(\nu(e))$, it must be the case that $w \notin \mathcal{L}((\Sigma^* b \cup \epsilon)\nu(f)b\Sigma^*)$. It follows that $w_1 \notin \mathcal{L}((\Sigma^* b \cup \epsilon)\nu(f))$. But since $w_1$ is of the form $u1^{2i+1}1a1^{2i+2}1a$, it follows that $u = \epsilon$ or $u = u'b$, for some $u' \in \Sigma^*$. In any case it must hold that $1^{2i+1}1a1^{2i+2}1a \notin \mathcal{L}(\nu(f))$. Notice, however, that $\mathcal{L}(1^{2i+1}\nu(x_i)a1^{2i+2}\nu(\bar{x}_i)a) \subseteq \mathcal{L}(\nu(f))$. Hence, $1^{2i+1}1a1^{2i+2}1a \in \mathcal{L}(\nu(f))$, which is a contradiction. Suppose, on the other hand, that $\sigma(x_i) = \sigma(\bar{x}_i) = 0$. Consider the unique prefix $w_1$ of $w$ that is of the form $u1^{2i+1}0a1^{2i+2}0a$, for $u \in \Sigma^*$. Then $w$ is of the form $w_1 w_2$, where $w_2 \in b\Sigma^*$. Since $w \notin \mathcal{L}(\nu(e))$, it must be the case that $w \notin \mathcal{L}((\Sigma^* b \cup \epsilon)\nu(f)b\Sigma^*)$. It follows that $w_1 \notin \mathcal{L}((\Sigma^* b \cup \epsilon)\nu(f))$. But since $w_1$ is of the form $u1^{2i+1}0a1^{2i+2}0a$, it follows that $u = \epsilon$ or $u = u'b$, for some $u' \in \Sigma^*$. In any case it must hold that $1^{2i+1}0a1^{2i+2}0a \notin \mathcal{L}(\nu(f))$. Notice, however, that $\mathcal{L}(1^{2i+1}\nu(x_i)a1^{2i+2}\nu(\bar{x}_i)a) \subseteq \mathcal{L}(\nu(f))$. Hence, $1^{2i+1}0a1^{2i+2}0a \in \mathcal{L}(\nu(f))$, which is a contradiction.

We can then define a propositional assignment $\sigma : \{p_1, \ldots, p_m\} \to \{0, 1\}$ such that $\sigma(p_i) = \nu(x_i)$, for each $1 \leq i \leq m$. Notice, from our previous remarks, that $\sigma(\neg p_i) = 1 - \nu(x_i) = \nu(\bar{x}_i)$, for each $1 \leq i \leq m$. We prove next that $\sigma$ satisfies $\varphi$. Assume this is not the case. Then for some $1 \leq j \leq n$ it is the case that $\sigma(\ell_j^1) = \sigma(\ell_j^2) = \sigma(\ell_j^3) = 0$. Consider now the unique prefix $w_1$ of $w$ such that $w_1$ is of the form $ub1^{3j(m+1)+1}0a1^{3j(m+1)+2}0a1^{3j(m+1)+3}0a$, for $u \in \Sigma^*$. Then $w$ is of the form $w_1 w_2$, where $w_2 \in b\Sigma^*$. Since $w \notin \mathcal{L}(\nu(e))$, it must be the case that $w \notin \mathcal{L}(\Sigma^* b\nu(f)b\Sigma^*)$. It follows that $w_1 \notin \mathcal{L}(\Sigma^* b\nu(f))$. But since $w_1$ is of the form $ub1^{3j(m+1)+1}0a1^{3j(m+1)+2}0a1^{3j(m+1)+3}0a$, it follows that $1^{3j(m+1)+1}0a1^{3j(m+1)+2}0a1^{3j(m+1)+3}0a \notin \mathcal{L}(\nu(f))$. Let $q_1$, $q_2$ and $q_3$ be the variables in $e$ associated with $\ell_j^1$, $\ell_j^2$ and $\ell_j^3$, respectively. Then it cannot be the case that $\nu(q_1) = \nu(q_2) = \nu(q_3) = 0$. Assume otherwise. It is clear that $\mathcal{L}(1^{3j(m+1)+1}\nu(q_1)a1^{3j(m+1)+2}\nu(q_2)a1^{3j(m+1)+3}\nu(q_3)a) \subseteq \mathcal{L}(\nu(f))$, and, therefore, $1^{3j(m+1)+1}0a1^{3j(m+1)+2}0a1^{3j(m+1)+3}0a \in \mathcal{L}(\nu(f))$, which is a contradiction. Thus, either $\nu(q_1) = \sigma(\ell_j^1) = 1$ or $\nu(q_2) = \sigma(\ell_j^2) = 1$ or $\nu(q_3) = \sigma(\ell_j^3) = 1$. This is our desired contradiction.

We prove second that if $\varphi$ is satisfiable then $w \notin \mathcal{L}_\square(e)$. Assume that $\varphi$ is satisfiable. Then there exists a propositional assignment $\sigma : \{p_1, \ldots, p_m\} \to \{0, 1\}$ that satisfies $\varphi$. We define a valuation $\nu : \{x_1, \bar{x}_1, \ldots, x_m, \bar{x}_m\} \to \{0, 1\}$ for $e$ as follows: For each $1 \leq i \leq m$ it is the case that $\nu(x_i) = \sigma(p_i)$ and $\nu(\bar{x}_i) = 1 - \sigma(p_i)$. We prove next that $w \notin \mathcal{L}(\nu(e))$.

Clearly, $w \notin \mathcal{L}(\nu(e))$ if and only if for each words $w_1, w_2, w_3 \in \Sigma^*$ such that $w = w_1 w_2 w_3$ it is the case that $w_1 \notin \mathcal{L}(\Sigma^* b \cup \epsilon)$ or $w_2 \notin \mathcal{L}(\nu(f))$ or $w_3 \notin \mathcal{L}(b\Sigma^* \cup \epsilon)$. Thus, in order to prove that $w \notin \mathcal{L}(\nu(e))$ it is enough to prove that for each words $w_1, w_2, w_3 \in \Sigma^*$ such that $w = w_1 w_2 w_3$,

(*)    if $w_1 \in \mathcal{L}(\Sigma^* b \cup \epsilon)$ and $w_3 \in \mathcal{L}(b\Sigma^* \cup \epsilon)$ then $w_2 \notin \mathcal{L}(\nu(f))$.

Take arbitrary words $w_1, w_2, w_3 \in \Sigma^*$ such that $w = w_1 w_2 w_3$. We consider several cases:

1. Either $w_1 \notin \mathcal{L}(\Sigma^* b \cup \epsilon)$ or $w_3 \notin \mathcal{L}(b\Sigma^* \cup \epsilon)$. Then (*) is trivially true.

2. It is the case that $w_1 \in \mathcal{L}(\Sigma^* b \cup \epsilon)$, $w_3 \in \mathcal{L}(b\Sigma^* \cup \epsilon)$, and $w_2$ is of the form $1^{2i+1}1a1^{2i+2}1au$, for some $1 \leq i \leq m$ and $u \in \Sigma^*$. Assume, for the sake of contradiction, that $w_2 \in \mathcal{L}(\nu(f))$. Since clearly there is no word accepted by $\mathcal{L}(\nu(f))$ with prefix $baa$, it must be the case that $w_3$ is not the empty word, and, therefore, that $w_3 \in \mathcal{L}(\Sigma^* b)$. Thus, the only possibility for $w_2$ to belong to $\mathcal{L}(\nu(f))$ is that $1^{2i+1}1a \in \mathcal{L}(\nu(f_i))$ and $1^{2i+2}1a \in \mathcal{L}(\nu(g_i))$. But this can only happen if $\nu(x_i) = 1$ and $\nu(\bar{x}_i) = 1$, which is our desired contradiction (since $\nu(x_i) = 1 - \nu(\bar{x}_i)$).

3. It holds that $w_1 \in \mathcal{L}(\Sigma^* b \cup \epsilon)$, $w_3 \in \mathcal{L}(b\Sigma^* \cup \epsilon)$, and $w_2$ is of the form $1^{2i+1}0a1^{2i+2}0au$, for some $1 \leq i \leq m$ and $u \in \Sigma^*$. This case is completely analogous to the previous one.

4. It is the case that $w_1 \in \mathcal{L}(\Sigma^* b \cup \epsilon)$, $w_3 \in \mathcal{L}(b\Sigma^* \cup \epsilon)$, and $w_2$ is of the form $1^{3j(m+1)+1}0a1^{3j(m+1)+2}0a1^{3j(m+1)+3}0au$, for some $1 \leq i \leq m$ and $u \in \Sigma^*$. Assume, for the sake of contradiction, that $w_2 \in \mathcal{L}(\nu(f))$. It is easy to see that the only way in which this can happen is that $\nu(q_1) = \nu(q_2) = \nu(q_3) = 0$, where $q_1$, $q_2$ and $q_3$ are the variables in $e$ that are associated with $\ell_j^1$, $\ell_j^2$ and $\ell_j^3$, respectively. Thus, $\sigma(\ell_j^1) = \sigma(\ell_j^2) = \sigma(\ell_j^3) = 0$, which is or desired contradiction.

This finishes the proof of the proposition.

### Proof of Theorem 3:

The hardness is established via a reduction from the complement 3-SAT, that is based on the $\Sigma_2^P$-hardness proof of Proposition 1.

Let $\varphi := C_1 \wedge \cdots \wedge C_p$ be an instance of 3-SAT, that uses variables $\{y_1, \ldots, y_t\}$.

Let $w$ be the word 0011000. From $\varphi$ we construct in polynomial time a parameterized regular expression $e$ over alphabet $\Sigma = \{0, 1\}$ such that $\varphi$ is not satisfiable if and only if $\mathcal{L}_\square(e)$ contains $w$.

Assume that each $C_j$ ($1 \leq j \leq p$) is of form $(\ell_j^1 \vee \ell_j^2 \vee \ell_j^3)$, where each literal $\ell_j^i$, for $1 \leq j \leq p$ and $1 \leq i \leq 3$, is either a variable in $\{y_1, \ldots, y_t\}$, or its negation.

With each propositional variable $y_k$, $1 \leq k \leq t$, we associate fresh variables $Y_k$ and $\hat{Y}_k$. Let $h$ be a function that maps each literal $\ell_j^i$ to the variable $Y_k$, if $\ell_j^i$ corresponds to $y_k$ or to $\hat{Y}_k$, if $\ell_j^i$ corresponds to $\neg y_k$. Expression $e$ is defined as

$$e = e_1 \mid e_{2,1} \mid \cdots \mid e_{2,t},$$

where $e_1 = 0011\big(h(\ell_1^1) \cdot h(\ell_1^2) \cdot h(\ell_1^3) \mid \cdots \mid h(\ell_p^1) \cdot h(\ell_p^2) \cdot h(\ell_p^3)\big)$, and for each $1 \leq k \leq t$, $e_{2,k} = \big(Y_k \hat{Y}_k 11000\big) \mid \big(00 Y_k \hat{Y}_k 000\big)$.

The proof that $w \in \mathcal{L}_\square(e)$ if and only if $\varphi$ is not satisfiable goes along the same lines as the $\Sigma_2^P$ hardness proof of Proposition 1.

Second, we prove that, for each word $w \in \Sigma^*$, the problem MEMBERSHIP$_\square(w)$ can be solved in polynomial time (actually, in linear time with respect to the size of the expression).

In order to do this, we first define a high-level procedure CheckSimpleMemb$_\square$, that takes as input a simple parameterized regular expression $e$ over $\Sigma$ and a finite set $\mathcal{W} \subset \Sigma^*$, and checks whether there exists an assignment $\nu$ for $e$ such that no word from $\mathcal{W}$ belongs to $\mathcal{L}(\nu(e))$. Then the answer to MEMBERSHIP$_\square(w)$ for an expression $e$ is $\neg$CheckSimpleMemb$_\square(e, \{w\})$.

The procedure CheckSimpleMemb$_\square$ works recursively on input $e$ and $\mathcal{W}$. For each internal node of the parse tree of $e$ it iterates over some sets $\mathcal{W}_1$ (or pairs of sets $(\mathcal{W}_1, \mathcal{W}_2)$ respectively), and for each such set (or pair) calls itself recursively on the children of the analyzed node. If the returned answers in case of one of the sets (or pairs) satisfy a given condition, the call accepts.

The details of the definition of CheckSimpleMemb$_\square$ are following:

1. If $e = a$, for $a \in \Sigma$, then CheckSimpleMemb$_\square(e, \mathcal{W})$ accepts iff $a \notin \mathcal{W}$.

2. If $e = x$, for $x \in \mathcal{V}$, then CheckSimpleMemb$_\square(e, \mathcal{W})$ accepts iff $\mathcal{W}$ does not contain all one-letter words.

3. If $e$ is of the form $e_1 \cup e_2$, then CheckSimpleMemb$_\square(e, \mathcal{W})$ accepts iff CheckSimpleMemb$_\square(e_1, \mathcal{W})$ accepts and CheckSimpleMemb$_\square(e_2, \mathcal{W})$ accepts.

4. If $e$ is of the form $e_1 e_2$, then CheckSimpleMemb$_\square(e, \mathcal{W})$ accepts iff there exist sets $\mathcal{W}_1 \subset \Sigma^*$, $\mathcal{W}_2 \subset \Sigma^*$ such that: (1) For each word $w_1 w_2 \in \mathcal{W}$ either $w_1 \in \mathcal{W}_1$ or $w_2 \in \mathcal{W}_2$, and (2) CheckSimpleMemb$_\square(e_1, \mathcal{W}_1)$ accepts and (3) CheckSimpleMemb$_\square(e_2, \mathcal{W}_2)$ accepts.

5. If $e$ is of the form $(e_1)^*$, then CheckSimpleMemb$_\square(e, \mathcal{W})$ accepts iff there is a set $\mathcal{W}_1 \subset \Sigma^*$ such that: (1) For each word $w_1 w_2 \cdots w_k \in \mathcal{W}$ at least one $w_i$ ($1 \leq i \leq k$) belongs to $\mathcal{W}_1$, and (2) CheckSimpleMemb$_\square(e_1, \mathcal{W}_1)$ accepts.

It is good to see, why CheckSimpleMemb$_\square$ needs to operate on sets of words instead of single words. The above procedure may construct non-singleton sets in case of concatenation and Kleene star and we cannot analyse their elements separately, because in each case we must judge existence of a valuation $\nu$, which would simultaneously prevent *all* possible runs of $w$ on $\nu(e)$ from being accepting.

Now, we prove that the procedure descibed above is sound and complete; that is, we prove that for each simple expression $e$ over $\Sigma$ and $\mathcal{W} \subset \Sigma^*$, CheckSimpleMemb$_\square$ accepts input $e$ and $\mathcal{W}$ iff there exists a valuation $\nu$ for $e$ such that no word in $\mathcal{W}$ belongs to $\mathcal{L}(\nu(e))$. We do this by induction:

1. The basis cases – when $e = a$, for $a \in \Sigma$, or $e = x$, for $x \in \mathcal{V}$ – are trivial.

2. Assume $e$ is of the form $e_1 \cup e_2$. Then there is a valuation $\nu$ for $e$ such that no word in $\mathcal{W}$ belongs to $\mathcal{L}(\nu(e))$ iff there is a valuation $\nu$ for $e$ such that for each $w \in \mathcal{W}$ we have $w \notin \mathcal{L}(\nu(e_1))$ and $w \notin \mathcal{L}(\nu(e_2))$. But since we consider only simple expressions here, the latter holds iff there are valuations $\nu_1$ for $e_1$ and $\nu_2$ for $e_2$ such that (a) no word $w \in \mathcal{W}$ belongs to $\mathcal{L}(\nu_1(e_1))$, and (b) no word $w \in \mathcal{W}$ belongs to $\mathcal{L}(\nu_2(e_2))$. By induction hypothesis, the latter holds iff $\texttt{CheckSimpleMemb}_\square(e_1, \mathcal{W})$ accepts and $\texttt{CheckSimpleMemb}_\square(e_2, \mathcal{W})$ accepts, which, by definition, is equivalent to the fact that $\texttt{CheckSimpleMemb}_\square(e, \mathcal{W})$ accepts.

3. Assume $e$ is of the form $e_1 e_2$. Then there is a valuation $\nu$ for $e$ such that no word $w \in \mathcal{W}$ belongs to $\mathcal{L}(\nu(e))$ iff there is a valuation $\nu$ for $e$ such that for each word $w_1 w_2 \in \mathcal{W}$ either $w_1 \notin \mathcal{L}(\nu(e_1))$ or $w_2 \notin \mathcal{L}(\nu(e_2))$. But since we consider only simple expressions here, the latter holds iff there are valuations $\nu_1$ for $e_1$ and $\nu_2$ for $e_2$ such that for each $w_1 w_2 \in \mathcal{W}$ either $w_1 \notin \mathcal{L}(\nu_1(e_1))$ or $w_2 \notin \mathcal{L}(\nu_2(e_2))$.

   Clearly, the latter holds iff there are valuations $\nu_1$ for $e_1$ and $\nu_2$ for $e_2$ and there are finite sets $\mathcal{W}_1, \mathcal{W}_2 \subset \Sigma^*$ such that: (1) For each $w_1 w_2 \in \mathcal{W}$ either $w_1 \in \mathcal{W}_1$ or $w_2 \in \mathcal{W}_2$, and (2) no word $w_1 \in \mathcal{W}_1$ belongs to $\mathcal{L}(\nu_1(e_1))$, and (3) no word $w_2 \in \mathcal{W}_2$ belongs to $\mathcal{L}(\nu_2(e_2))$. By induction hypothesis, the latter holds iff there are finite sets $\mathcal{W}_1, \mathcal{W}_2 \subset \Sigma^*$ such that for each word $w_1 w_2 \in \mathcal{W}$ either $w_1 \in \mathcal{W}_1$ or $w_2 \in \mathcal{W}_2$, and both $\texttt{CheckSimpleMemb}_\square(e_1, \mathcal{W}_1)$ and $\texttt{CheckSimpleMemb}_\square(e_2, \mathcal{W}_2)$ accept. By definition, the latter is equivalent to the fact that $\texttt{CheckSimpleMemb}_\square(e, \mathcal{W})$ accepts.

4. Assume $e$ is of the form $(e_1)^*$. Then there is a valuation $\nu$ for $e$ such that no word $w \in \mathcal{W}$ belongs to $\mathcal{L}(\nu(e))$ iff there is a valuation $\nu_1$ for $e_1$ such that for each $w_1 w_2 \ldots w_k \in \mathcal{W}$ some $w_i$ ($1 \le i \le k$) does not belong to $\mathcal{L}(\nu_1(e_1))$. Clearly, the latter holds iff there is a valuation $\nu_1$ for $e_1$ and there is a finite set $\mathcal{W}_1 \subset \Sigma^*$ such that: (1) For each word $w \in \mathcal{W}$ and each its decomposition $w = w_1 w_2 \ldots w_k$ some $w_i$ ($1 \le i \le k$) belongs to $\mathcal{W}_1$, and (2) no word from $\mathcal{W}_1$ belongs to $\mathcal{L}(\nu_1(e_1))$. By induction hypothesis, the latter holds iff $\texttt{CheckSimpleMemb}_\square(e_1, \mathcal{W}_1)$ accepts for some set $\mathcal{W}_1 \subset \Sigma^*$ satisfying condition (1), which, by definition, is equivalent to the fact that $\texttt{CheckSimpleMemb}_\square(e, \mathcal{W})$ accepts.

Next we show that there is an implementation of the procedure $\texttt{CheckSimpleMemb}_\square$ that works in $O(|e|)$ time, if we assume that the input consists of a simple parameterized regular expression $e$ and a *fixed* set of words $\mathcal{W}$.

First, for technical reasons, we remove all subexpressions $\epsilon$ from $e$ obtaining $e'$. Then the implementation works recursively as follows: If $e'$ is of the form $a$, for $a \in \Sigma$, or $x \in \mathcal{V}$, or $e_1 \cup e_2$, then we implement recursively in the same way as it is described in $\texttt{CheckSimpleMemb}_\square$. If, on the other hand, $e'$ is of the form $e_1 e_2$ or $(e_1)^*$, then we have to be slightly more careful since we have to define how to search for sets $\mathcal{W}_1$ and $\mathcal{W}_2$. We do this as follows:

1. Assume first that $e$ is of the form $e_1 e_2$. Then $\texttt{CheckSimpleMemb}_\square$ accepts $e'$ and $\mathcal{W}$ iff there are sets $\mathcal{W}_1, \mathcal{W}_2 \subset \Sigma^*$ such that: (1) If $w_1 w_2 \in \mathcal{W}$, then $w_1 \in \mathcal{W}_1$ or $w_2 \in \mathcal{W}_2$, and (2) $\texttt{CheckSimpleMemb}_\square(e_1, \mathcal{W}_1)$ accepts, and (3) $\texttt{CheckSimpleMemb}_\square(e_2, \mathcal{W}_2)$ accepts. Our implementation, however, does not look over arbitrary sets $\mathcal{W}_1$ and $\mathcal{W}_2$, but only over the sets which can be constructed as follows: For each $w \in \mathcal{W}$ and for each $w_1, w_2 \in \Sigma^+$ (that is, both $w_1$ and $w_2$ are nonempty) such that $w = w_1 w_2$, either pick up $w_1$ and place it in $\mathcal{W}_1$ or pick up $w_2$ and place it in $\mathcal{W}_2$. If for some pair $(\mathcal{W}_1, \mathcal{W}_2)$ constructed in this way it is the case that $\texttt{CheckSimpleMemb}_\square(e_1, \mathcal{W}_1)$ accepts and $\texttt{CheckSimpleMemb}_\square(e_2, \mathcal{W}_2)$ accepts, then $\texttt{CheckSimpleMemb}_\square(e', \mathcal{W})$ accepts. The reason why we can restrict ourselves to the case of nonempty words is that neither $e_1$ nor $e_2$ equals $\epsilon$ in $e'$, and, thus, the empty word trivially does not match any of them. Clearly our implementation continues being sound and complete.

2. Assume second that $e'$ is of the form $(e_1)^*$. Then $\texttt{CheckSimpleMemb}_\square$ accepts iff there is a set $\mathcal{W}_1 \subset \Sigma^*$ such that: (1) For each word $w \in \mathcal{W}$ and each its decomposition $w = w_1 w_2 \ldots w_k$ some $w_i$ $(1 \leq i \leq k)$ belongs to $\mathcal{W}_1$, and (2) $\texttt{CheckSimpleMemb}_\square(e_1, \mathcal{W}_1)$ accepts. Again, our implementation does not look over arbitrary sets $\mathcal{W}_1$, but only over the sets which can be constructed as follows: For each decomposition $w_1 w_2 \cdots w_k$ of each word in $\mathcal{W}$ pick up an arbitrary $1 \leq i \leq k$ and place $w_i$ in $\mathcal{W}_1$. If for some set $\mathcal{W}_1$ constructed in this way $\texttt{CheckSimpleMemb}_\square(e_1, \mathcal{W}_1)$ accepts, then $\texttt{CheckSimpleMemb}_\square(e', \mathcal{W})$ accepts. Clearly, our implementation continues being sound and complete.

Too estimate the time complexity of the above implementation, first we need to see that all elements of all $\mathcal{W}$ encountered in the algorithm are subwords of $w$ and thus all $\mathcal{W} \in \mathbb{W}$, where $\mathbb{W}$ is the powerset of all subwords of $w$. Clearly the size of $\mathbb{W}$ is dependent only on $|w|$. Also the number of cases tried by each subcall of $\texttt{CheckSimpleMemb}_\square$ and the number of steps needed to construct each $\mathcal{W}_1$ (and $\mathcal{W}_2$ respectively) is dependent only on $|w|$. Hence all these values are constant.

If along the algorithm we memoize the answers to subcalls, then our complexity will be upper-bounded by the complexity of a dynamic version of the above algorithm, which would calculate $\texttt{CheckSimpleMemb}_\square(e_1, \mathcal{W}_1)$ for all subexpressions $e_1$ of $e'$ and all $\mathcal{W}_1 \in \mathbb{W}$ in a bottom-up order. In this approach, the computation of $\texttt{CheckSimpleMemb}_\square(e_1, \mathcal{W}_1)$ would take constant time, because answers to subcalls would have been precomputed. Thus the total complexity of $\texttt{CheckSimpleMemb}_\square$ is linear with respect to the size of the parse tree of $e'$ and thus with respect to $e$ as well.

We can strive for accelerating the above algorithm, by replacing all subexpressions of the form $(f_1 \cup \cdots \cup f_i^* \cup \cdots f_n)^*$ with $(f_1 \cup \cdots \cup f_i \cup \cdots f_n)^*$, which does not change the semantics of the expression, but avoids unnecessary computational effort. It will not lower the asymptotic complexity, but will help lower the constant.

We finally prove that, for each word $w \in \Sigma^*$, the problem $\text{MEMBERSHIP}_\diamond(w)$ can be solved in time $O(n \log^2 n)$. Obviously, we can concern all labels $a \in \Sigma$, which do not occur in $w$ as equal. This simple observation makes the alphabet size fixed: $|\Sigma| \leq |w| + 1$. Now, let $w$ be a word over $\Sigma$. Next we construct an algorithm that, given a parameterized regular expression $e$ over $\Sigma$, checks whether $w \in \mathcal{L}_\diamond(e)$. Using techniques from [13] the algorithm first constructs an NFA $\mathcal{A}$ over $\Sigma \cup \mathcal{W}$ that is equivalent to $e$, with $O(n)$ states and $O(n \log^2 n)$ transitions, and then performs a nondeterministic logarithmic space algorithm on $\mathcal{A}$.

Assume that $\mathcal{W} \subset \mathcal{V}$ is the set of variables that appear in $e$. Construct, in polynomial time, an NFA $\mathcal{A}$ (with set of states $Q$) over $\Sigma \cup \mathcal{W}$ such that $\mathcal{L}(\mathcal{A}) = \mathcal{L}(e)$. Let us assume, without loss of generality, that $q_0$ is the unique initial state of $\mathcal{A}$. Further, assume that $w = a_1 a_2 \cdots a_m$, where each $a_i$ $(1 \leq i \leq m)$ is a symbol in $\Sigma$. Then we perform the following nondeterministic algorithm over $\mathcal{A}$: The algorithm works in at most $m+1$ steps. At each step $0 \leq i \leq m$ the state of the algorithm consists of a pair $(q_i, \mu_i)$, where $q_i \in Q$ and $\mu_i$ is a mapping from some subset $\mathcal{W}_i$ of $\mathcal{W}$ into $\Sigma$. The initial state of the algorithm is $(q_0, \mu_0)$, where $\mu_0 : \emptyset \to \Sigma$ (recall that $q_0$ is the initial state of $\mathcal{A}$). Assume that the state of the algorithm in step $i < m$ is $(q_i, \mu_i)$. Then in step $i+1$ the algorithm nondeterministically picks up a pair $(q_{i+1}, \mu_{i+1})$ and checks that at least one of the following conditions holds:

- There exists a transition labeled $a_i \in \Sigma$ from $q_i$ to $q_{i+1}$ in $\mathcal{A}$ and $\mu_i = \mu_{i+1}$; that is, both $\mu_i$ and $\mu_{i+1}$ are mappings from $\mathcal{W}_i$ into $\Sigma$, and $\mu_{i+1}(x) = \mu_i(x)$, for each $x \in \mathcal{W}_i$.

- There exists a transition labeled $x \in \mathcal{V}$ from $q_i$ to $q_{i+1}$ in $\mathcal{A}$, $x \notin \mathcal{W}_i$ and $\mu_{i+1} : \mathcal{W}_i \cup \{x\} \to \Sigma$ is defined as follows: $\mu_{i+1}(y) = \mu_i(y)$, for each $y \in \mathcal{W}_i$, and $\mu_{i+1}(x) = a_i$.

- There exists a transition labeled $x \in \mathcal{V}$ from $q_i$ to $q_{i+1}$ in $\mathcal{A}$, $x \in \mathcal{W}_i$, $\mu_i(x) = a_i$ and $\mu_i = \mu_{i+1}$.

The procedure accepts if it reaches step $n$ in state $(q_n, \mu_n)$, for some accepting state $q_n$ of $\mathcal{A}$. Notice that, since $w$ is fixed, the size of each mapping $\mu$ from a subset of $\mathcal{W}$ into $\Sigma$ is also fixed: $|\mu| \leq \min\{|w|, |\mathcal{W}|\}$. That is because the initial mapping is empty and in each step of the algorithm it can grow only by one. This means that the nondeterministic procedure described above works in NLOGSPACE.

It is not hard to prove (esentially using the same techniques than in the second part of the proof of Proposition 2) that the procedure described above accepts the parameterized regular expression $e$ if and only if $w \in \mathcal{L}_\diamond(e)$.

Now let $M$ be the set of all mappings $\mu$ from subsets of $\mathcal{W}$ to $\Sigma$, which can occur in the algorithm presented above, and $V$ be the number of all states $(q, \mu)$, which can occur therein. To see the precise time complexity, we need to estimate $|M|$ and $|V|$. First, $|M| = O\left(|\Sigma|^{\min\{|w|, |\mathcal{W}|\}}\right)$, which is fixed in our case. Then, $V = O(n \cdot |M|)$, which is linear in the number of states of $\mathcal{A}$.

Now let us imagine a directed graph $G$ with the set of vertices $V$, in which there is an edge from state $(q, \mu)$ to state $(q', \mu')$ iff the pair $(q', \mu')$ can be picked up from pair $(q, \mu)$ according to the algorithm presented above. Each edge $(q, \mu) \to (q', \mu')$ corresponds to an edge $q \to q'$ in $\mathcal{A}$ and for each edge $q \to q'$ in $\mathcal{A}$ there are at most $|M|$ edges $(q, \mu) \to (q', \mu')$ (one for each $\mu \in M$). Therefore, $G$ has $O(n \log^2 n)$ edges, since $|M|$ is fixed. We can also construct $G$ in $O(n \log^2 n)$ time and space.

Finally, it suffices to perform a reachability search in $G$ to see whether an accepting state can be reached from node $(q_0, \mu_0)$ in $G$, which can clearly be done in linear time with respect to the size of $G$, which gives us an algorithm with $O(n \log^2 n)$ time complexity or, by dropping the assumption of $w$ being fixed — $O\left(|w| \cdot |\Sigma|^{\min\{|w|, |\mathcal{W}|\}} \cdot n \log^2 n\right)$ time complexity. Moreover, we can spare space by not constructing $G$ and by computing it "on the fly", because standard graph search algorithms run in $O(V)$ space. It might be also useful in terms of time, because a fixed word $w$ either attains an accepting state within a short path or does not do so at all, so usually most part of $G$ would not by touched by the search algorithm at all.

It is worth analysing the gain in performance, which the above method gives in comparison to the direct approach. The straightforward algorithm calculates an NFA accepting $\mathcal{L}_\diamond(e)$ and runs reachability search on it. The size of such NFA is $O(|\Sigma|^{|\mathcal{W}|})$, so the time complexity becomes $O(|w| \cdot |\Sigma|^{|\mathcal{W}|} \cdot n \log^2 n)$. Hence, the only gain, that the former algorithm gives is lowering the exponent over $\Sigma$ from $|\mathcal{W}|$ to $\min\{|w|, |\mathcal{W}|\}$ and this is because we it takes into advantage a smaller class of mappings, confined by the length of the run of $w$ on $\mathcal{A}$. In fact, this gain is very large if we speak of problem instances with a relatively small $w$ and a huge $e$.

## Proof of Theorem 4

(Part 1) We begin with the proof for the first part. The lower bound follows directly from the fact that checking universality is PSPACE-hard even for complete regular expressions. For the upper bound, we need the following easy claim from the definition of certain acceptance:

**Claim 2** *Let $e$ be a parameterized regular expression over an alphabet $\Sigma$. Then, $\mathcal{L}_\Box(e) = \Sigma^*$ if and only if $L(v(e)) = \Sigma^*$, for every valuation $v$ for $e$.*

Then, a PSPACE algorithm to solve the complement of the UNIVERSALITY$_\Box$ problem (wether $\mathcal{L}_\Box(e) \neq \Sigma^*$) just guesses a valuation $v$ for $e$, and then checks wether $L(v(e)) \neq \Sigma^*$. Since $v(e)$ is a complete regular expression, it is well known that this decision procedure can be performed in PSPACE. This finishes the proof of the first part of the theorem.

(Part 2) Next, we prove the second part, that is, we show that UNIVERSALITY$_\Diamond$ is EXPSPACE-complete. We begin with the upper bound. Given a parameterized regular expression $e$, it is easy to see that an equivalent, complete regular expression $e'$ such that $\mathcal{L}_\Diamond(e) = \mathcal{L}(e')$ can be constructed in exponential time: just take $\bigcup_{v \text{ is a valuation for } e} v(e)$ (the number of possible valuations is $|\Sigma|^{|\mathcal{W}|}$). Combining this fact with the well known result that there is an algorithm to check wether the language of a (complete) regular expression $e'$ is universal that require only polynomial space w.r.t. $e'$, we obtain our EXPSPACE algorithm: First obtain a complete regular expression $e'$ such that $\mathcal{L}_\Diamond(e) = \mathcal{L}(e')$, and then decide wether $\mathcal{L}(e') = \Sigma^*$.

For the upper bound we present a reduction from the complement of the acceptance problem of a Turing machine. Let $L$ be a language that belongs to EXPSPACE, and let $\mathcal{M}$ be a Turing machine that decides $L$ in EXPSPACE. Given an input $\bar{a} = a_0, \ldots, a_{k-1}$, we construct in polynomial time with respect to $\mathcal{M}$ and $\bar{a}$ a parameterized regular expression $e_{\mathcal{M},\bar{a}}$ such that $\mathcal{L}_\Diamond(e_{\mathcal{M},\bar{a}}) = \Sigma^*$ if and only if $\mathcal{M}$ does not accepts $\bar{a}$.

Assume that $\mathcal{M} = \{Q, \Sigma, \Gamma, q_0, \{q_m\}, \delta\}$; that is, the states of $\mathcal{M}$ are $Q = \{q_0, \ldots, q_m\}$, the initial state is $s_0$, the input alphabet is $\Sigma$, and $\Gamma$ is the union of $\Sigma$ plus a number of symbols reserved for the Turing machine, and that the set of transitions of $\mathcal{M}$ is $\delta$. Without loss of generality, we assume that $\mathcal{M}$ has only one tape, starts with the input copied on the first $|\bar{a}|$ cells of this tape, has only one final state $s_m$, and that no transition is defined for that state. Moreover, since $\mathcal{M}$ decides $L$ in EXPSPACE, there is a polynomial $S()$ such that, for every input $\bar{a}$ over $\Sigma$, $\mathcal{M}$ decides $\bar{a}$ using space of order $2^{S(|\bar{a}|)}$. Assume for notation convenience that $S(|\bar{a}|) = n$, and as usual $\Gamma = \Sigma \cup \{B\}$.

Let $\Delta = \{0, 1, \$, \%, \#, \&\} \cup \Gamma \cup (\Gamma \times Q)$. Assume that $\Sigma = \{b_1, \ldots, b_p\}$. For the sake of readability, for a set $B = \{b_1, \ldots, b_n\}$ of symbols we denote by $B$ the regular expression $b_p \mid \cdots \mid b_p$. Thus, for example, assume that $\Gamma = \Sigma \cup \{B\}$. Then, when we write $(\Gamma \cup (\Gamma \times Q))$ we represent the language given by $(b_1 \mid \cdots \mid b_p \mid B \mid (b_1, s_0) \mid \cdots \mid (b_p, s_m))$. Using the alphabet $\Delta$, we represent a configuration of the Turing machine by a word in the language

$$\#(\$[0]\%(\Gamma \cup (\Gamma \times Q))) \cdots (\$[2^n - 1]\%(\Gamma \cup (\Gamma \times Q)))\&$$

Next we construct a parameterized regular expression $e_{\mathcal{M},\bar{a}}$ such that $\mathcal{L}_\Diamond(e) \neq \Sigma^*$ if and only if $\mathcal{M}$ accepts on input $\bar{a}$. Define $e_{\mathcal{M},\bar{a}} = e^{\text{form}} \mid e^i \mid e^f \mid e^{\text{trans}}$, where

- $e^{\text{form}}$ describes all the words that do not represent a concatenation of configurations of $\mathcal{M}$.

- $e^i$ describes words that do not start with the initial configuration of $\mathcal{M}$ over input $a$.

- $e^f$ describes words that do not end in a final configuration for $\mathcal{M}$
- $e^{\text{trans}}$ describes words that contains two consecutive configurations $\alpha$ and $\beta$ such that $\alpha$ and $\beta$ do not agree on $\delta$.

We now describe these expressions. Expression $e^{form}$ is the union of the following expressions, describing that:

- The first symbol of the word is not #:
$$e_1^{\text{form}} = (\epsilon \mid (\Delta - \{\#\}))\Delta^*$$

- The last symbol of the word is not &:
$$e_2^{\text{form}} = \Delta^*(\Delta - \{\&\})$$

- After a # we do not have the symbol \$:
$$e_3^{\text{form}} = \Delta^*\#(\Delta - \{\$\})\Delta^*$$

- Between the symbols \$ and % there are less than $n$ symbols:
$$e_4^{\text{form}} = \Delta^*\$(\epsilon \mid \Delta)^{n-1}\%\Delta^*$$

- Between the symbols \$ and % there are more than $n$ symbols:
$$e_5^{\text{form}} = \Delta^*\$(\Delta - \{\%\})^{n+1}(\Delta - \{\%\})^*\%\Delta^*$$

- Between the symbols \$ and % there is symbol not in $\{0, 1\}$:
$$e_6^{\text{form}} = \Delta^*\$(0 \mid 1)^*(\Delta - \{0, 1, \%\})\Delta^*$$

- After a word in $(0|1)^n$ we do not have the symbol %:
$$e_7^{\text{form}} = \Delta^*(0 \mid 1)^n(\Delta - \{\%\})\Delta^*$$

- After the symbol % we do not have a symbol in $(\Gamma \cup (\Gamma \times Q))$
$$e_8^{\text{form}} = \Delta^*\%(\Delta - (\Gamma \cup (\Gamma \times Q)))\Delta^*$$

- After a word in $[i]\%(\Gamma \cup (\Gamma \times Q))$ we do not have the symbol \$, for $0 \leq i \leq 2^n - 2$:
$$\begin{aligned}
e_{9_1}^{\text{form}} &= \Delta^*0(0 \mid 1)^{n-1}\%(\Gamma \cup (\Gamma \times Q))(\Delta - \{\$\})\Delta^* \\
e_{9_2}^{\text{form}} &= \Delta^*(0 \mid 1)0(0 \mid 1)^{n-2}\%(\Gamma \cup (\Gamma \times Q))(\Delta - \{\$\})\Delta^* \\
&\vdots = \vdots \\
e_{9_n}^{\text{form}} &= \Delta^*(0 \mid 1)^{n-1}0\%(\Gamma \cup (\Gamma \times Q))(\Delta - \{\$\})\Delta^*
\end{aligned}$$

- After a word in $[2^n - 1]\%(\Gamma \cup (\Gamma \times Q))$ we do not have the symbol &:

$$e_{10}^{\text{form}} = \Delta^* 1^n \%(\Gamma \cup (\Gamma \times Q))(\Delta - \{\&\})\Delta^*$$

- After the symbol & we do not have the symbol #:

$$e_{11}^{\text{form}} = \Delta^* \&(\Delta - \{\#\})\Delta^*$$

- Between a symbol # and & there is no symbol in $\Gamma \times Q$ (a configuration does not have a reading position):

$$e_{12}^{\text{form}} = \Delta^* \#(\Delta - \{\&\} - (\Gamma \times Q))^* \& \Delta^*$$

- Between a symbol # and & there is more than one symbol in $\Gamma \times Q$ (a configuration features two positions being read by the machine):

$$e_{13}^{\text{form}} = \Delta^* \#(\Delta - \{\&\})^*(\Gamma \times Q)(\Delta - \{\&\})^*(\Gamma \times Q)(\Delta - \{\&\})^* \& \Delta^*$$

- After the word #$ we do not have the word $[0]$:

$$\begin{aligned}
e_{14_1}^{\text{form}} &= \Delta^* \#\$ 1(0 \mid 1)^{n-1} \Delta^* \\
e_{14_2}^{\text{form}} &= \Delta^* \#\$(0 \mid 1)1(0 \mid 1)^{n-2} \Delta^* \\
\vdots &= \vdots \\
e_{14_n}^{\text{form}} &= \Delta^* \#\$(0 \mid 1)^{n-1} 1 \Delta^*
\end{aligned}$$

- After a word $\$[i]\%(\Gamma \cup (\Gamma \times Q))$ we do not follow with $[i+1]\%(\Gamma \cup (\Gamma \times Q))$, where $i$ is even:

$$\begin{aligned}
e_{15_1}^{\text{form}} &= \Delta^*(0 \mid 1)^{n-1} 0\%(\Gamma \cup (\Gamma \times Q))\$(0 \mid 1)^{n-1} 0 \Delta^* \\
e_{15_{2,1}}^{\text{form}} &= \Delta^*(0 \mid 1)^{n-2} 00\%(\Gamma \cup (\Gamma \times Q))\$(0 \mid 1)^{n-2} 11 \Delta^* \\
e_{15_{2,2}}^{\text{form}} &= \Delta^*(0 \mid 1)^{n-2} 10\%(\Gamma \cup (\Gamma \times Q))\$(0 \mid 1)^{n-2} 01 \Delta^* \\
e_{15_{3,1}}^{\text{form}} &= \Delta^*(0 \mid 1)^{n-3} 0(0 \mid 1)0\%(\Gamma \cup (\Gamma \times Q))\$(0 \mid 1)^{n-3} 1(0 \mid 1)1 \Delta^* \\
e_{15_{3,2}}^{\text{form}} &= \Delta^*(0 \mid 1)^{n-3} 1(0 \mid 1)0\%(\Gamma \cup (\Gamma \times Q))\$(0 \mid 1)^{n-3} 0(0 \mid 1)1 \Delta^* \\
\vdots &= \vdots \\
e_{15_{n,1}}^{\text{form}} &= \Delta^* 0(0 \mid 1)^{n-2} 0\%(\Gamma \cup (\Gamma \times Q))\$ 1(0 \mid 1)^{n-2} 1 \Delta^* \\
e_{15_{n,2}}^{\text{form}} &= \Delta^* 1(0 \mid 1)^{n-2} 0\%(\Gamma \cup (\Gamma \times Q))\$ 0(0 \mid 1)^{n-2} 1 \Delta^*
\end{aligned}$$

- After a word $\$[i]\%(\Gamma \cup (\Gamma \times Q))$ we do not follow with $[i+1]\%(\Gamma \cup (\Gamma \times Q))$, where $i$ is odd

and $i < 2^n - 1$:

$$
\begin{aligned}
e^{\text{form}}_{16_1} &= \Delta^*(0 \mid 1)^{n-1}1\%(\Gamma \cup (\Gamma \times Q))\$(0 \mid 1)^{n-1}1\Delta^* \\
e^{\text{form}}_{16_{2,1}} &= \Delta^*(0 \mid 1)^{n-2}01\%(\Gamma \cup (\Gamma \times Q))\$(0 \mid 1)^{n-2}00\Delta^* \\
e^{\text{form}}_{16_{2,2,1}} &= \Delta^*(0 \mid 1)^{n-3}001\%(\Gamma \cup (\Gamma \times Q))\$(0 \mid 1)^{n-3}110\Delta^* \\
e^{\text{form}}_{16_{2,2,2}} &= \Delta^*(0 \mid 1)^{n-3}101\%(\Gamma \cup (\Gamma \times Q))\$(0 \mid 1)^{n-3}010\Delta^* \\
e^{\text{form}}_{16_{2,3,1}} &= \Delta^*(0 \mid 1)^{n-4}0(0 \mid 1)01\%(\Gamma \cup (\Gamma \times Q))\$(0 \mid 1)^{n-4}1(0 \mid 1)10\Delta^* \\
e^{\text{form}}_{16_{2,3,2}} &= \Delta^*(0 \mid 1)^{n-4}1(0 \mid 1)01\%(\Gamma \cup (\Gamma \times Q))\$(0 \mid 1)^{n-4}0(0 \mid 1)10\Delta^* \\
\vdots &= \vdots \\
e^{\text{form}}_{16_{2,n-2,1}} &= \Delta^*0(0 \mid 1)^{n-3}01\%(\Gamma \cup (\Gamma \times Q))\$1(0 \mid 1)^{n-3}10\Delta^* \\
e^{\text{form}}_{16_{2,n-2,2}} &= \Delta^*1(0 \mid 1)^{n-3}01\%(\Gamma \cup (\Gamma \times Q))\$0(0 \mid 1)^{n-3}10\Delta^* \\
\\
e^{\text{form}}_{16_{3,1}} &= \Delta^*(0 \mid 1)^{n-3}011\%(\Gamma \cup (\Gamma \times Q))\$(0 \mid 1)^{n-3}(000 \mid 110 \mid 010)\Delta^* \\
e^{\text{form}}_{16_{3,2,1}} &= \Delta^*(0 \mid 1)^{n-4}0011\%(\Gamma \cup (\Gamma \times Q))\$(0 \mid 1)^{n-3}1100\Delta^* \\
e^{\text{form}}_{16_{3,2,2}} &= \Delta^*(0 \mid 1)^{n-4}1011\%(\Gamma \cup (\Gamma \times Q))\$(0 \mid 1)^{n-4}0100\Delta^* \\
e^{\text{form}}_{16_{3,3,1}} &= \Delta^*(0 \mid 1)^{n-5}0(0 \mid 1)011\%(\Gamma \cup (\Gamma \times Q))\$(0 \mid 1)^{n-5}1(0 \mid 1)100\Delta^* \\
e^{\text{form}}_{16_{3,3,2}} &= \Delta^*(0 \mid 1)^{n-5}1(0 \mid 1)011\%(\Gamma \cup (\Gamma \times Q))\$(0 \mid 1)^{n-5}0(0 \mid 1)100\Delta^* \\
\vdots &= \vdots \\
e^{\text{form}}_{16_{3,n-3,1}} &= \Delta^*0(0 \mid 1)^{n-4}011\%(\Gamma \cup (\Gamma \times Q))\$1(0 \mid 1)^{n-4}100\Delta^* \\
e^{\text{form}}_{17_{3,n-3,2}} &= \Delta^*1(0 \mid 1)^{n-4}011\%(\Gamma \cup (\Gamma \times Q))\$0(0 \mid 1)^{n-4}100\Delta^* \\
\\
e^{\text{form}}_{16_{n,1}} &= \Delta^*01^{n-1}\%(\Gamma \cup (\Gamma \times Q))\$(\{(0 \mid 1)^n\} - \{1^n\})\Delta^*
\end{aligned}
$$

Notice that expression $e^{\text{form}}$ is of polynomial size. In particular, the language $(\{(0 \mid 1)^n\} - \{1^n\})$ can be described with the expression

$$0(0 \mid 1)^{n-1} \mid (0 \mid 1)0(0 \mid 1)^{n-3} \mid \cdots \mid (0 \mid 1)^{n-1}0$$

.

Next, expression $e^i$ is the union of the following expressions, describing that:

- The first configuration does not contain the initial state:

$$e^i_1 = \#(\$(0 \mid 1)^n\%(\Gamma \cup (\Gamma \times (Q - q_0))))^*\&\Delta^*$$

- The first configuration does not contain the head in the initial position:

$$e^i_2 = \#\$0^n\%\Gamma\Delta^*$$

- The first configuration does not have the word $\bar{a}$ in its first $|\bar{a}| = k$ symbols:

$$e^i_{3,1} = \#\$0^n\%(\Gamma \times Q - \{(q_0, a_0)\})\Delta^*$$
$$\vdots = \vdots$$
$$e^i_{3,k} = \#\$[0]\%(\Gamma \cup (\Gamma \times Q))\$[1]\%(\Gamma \cup (\Gamma \times Q)) \cdots \$[k-1]\%(\Gamma \times Q - \{(a_{k-1})\})\Delta^*$$

- The rest of the symbols of the first configuration are not blank symbols:

$$e^i_4 = \#\$[0]\%(\Gamma \cup (\Gamma \times Q))\$[1]\%(\Gamma \cup (\Gamma \times Q)) \cdots \$[k-1]\%(\Delta - \{\&\})^*(\Gamma \cup (\Gamma \times Q) - \{B\})\Delta^*$$

Furthermore, expression $ef$ describes words whose final configuration does not end in a final state:

$$e^f = \Delta^*\#(\Delta - \&)^*((\Gamma \times Q) - \{(a, s_m) \mid a \in \Gamma\})(\Delta - \&)^*\&$$

Finally, expression $e^{trans}$ is the union of the following expressions, describing that:

- A cell not pointed by the head changed it's content:

$$e^{trans}_1 = \bigcup_{a \in \Gamma} \Delta^*\#(\$(0|1)^n\%(\Gamma \cup (\Gamma \times Q)))^*\$x_1 \cdots x_n\%a(\$(0|1)^n\%(\Gamma \cup (\Gamma \times Q)))^*\&$$
$$\#(\$(0|1)^n\%(\Gamma \cup (\Gamma \times Q)))^*\$x_1 \cdots x_n\%(\Gamma - \{a\} \cup (\Gamma \times Q) - (\{a\} \times Q))(\$(0|1)^n\%(\Gamma \cup (\Gamma \times Q)))^*\&\Delta^*$$

- A configuration that is not final features a pair in $Q \times \Sigma$ for which no transition is defined (the last # states the configuration is not final):

$$e^{trans}_2 = \bigcup_{a \in \Gamma, s \in Q \mid \delta(s,a) \text{ is not defined}} \Delta^*\#(\$(0|1)^n\%(\Gamma \cup (\Gamma \times Q)))^*$$
$$\$(0|1)^n\%(a, s)(\$(0|1)^n\%(\Gamma \cup (\Gamma \times Q)))^*\&\#\Delta^*$$

- The change of state does not agree with $\delta$:

$$e^{trans}_3 = \bigcup_{A \in \Gamma, s \in Q \mid \delta(s,a) = (s', a', \{\rightarrow, \leftarrow\})} \Delta^*\#(\Delta - \{\&\})^*(s, a)(\Delta - \{\&\})^*\&$$
$$\#(\Delta - \{\&\})^*(\Gamma \times (Q - \{s'\}))(\Delta - \{\&\})^*\&\Delta^*$$

- The symbol written in a given step does not agree with $\delta$:

$$e^{trans}_3 = \bigcup_{a \in \Gamma, s \in Q \mid \delta(s,a) = (s', a', \rightarrow)}$$
$$\Delta^*\#(\$(0|1)^n\%(\Gamma \cup (\Gamma \times Q)))^*\$y_1, \ldots, y_n\%(a, s)(\$(0|1)^n\%(\Gamma \cup (\Gamma \times Q)))^*\&$$
$$\#(\$(0|1)^n\%(\Gamma \cup (\Gamma \times Q)))^*\$y_1, \ldots, y_n\%(\Gamma - \{a'\})(\$(0|1)^n\%(\Gamma \cup (\Gamma \times Q)))^*\&\Delta^*$$

- The movement of the head does not agree with $\delta$:

$$e_{4,1}^{\text{trans}} = \bigcup_{a \in \Gamma, s \in Q | \delta(s,a)=(s',a',\rightarrow)}$$
$$\Delta^* \#(\$(0|1)^n\%(\Gamma \cup (\Gamma \times Q)))^* \$x_1, \ldots, x_n\%(a,s)(\$(0|1)^n\%(\Gamma \cup (\Gamma \times Q)))^* \&$$
$$\#(\$(0|1)^n\%(\Gamma \cup (\Gamma \times Q)))^* \$x_1, \ldots, x_n\%(a')\epsilon | (\$(0|1)^n\%\Gamma(\$(0|1)^n\%(\Gamma \cup (\Gamma \times Q)))^*) \& \Delta^*$$

$$e_{4,2}^{\text{trans}} = \bigcup_{a \in \Gamma, s \in Q | \delta(s,a)=(s',a',\leftarrow)}$$
$$\Delta^* \#(\$(0|1)^n\%(\Gamma \cup (\Gamma \times Q)))^* \$x_1, \ldots, x_n\%(a,s)(\$(0|1)^n\%(\Gamma \cup (\Gamma \times Q)))^* \&$$
$$\#\epsilon | ((\$(0|1)^n\%(\Gamma \cup (\Gamma \times Q)))^* \$(0|1)^n\%\Gamma) \$x_1, \ldots, x_n\%(a')(\$(0|1)^n\%(\Gamma \cup (\Gamma \times Q)))^* \& \Delta^*$$

It is now straightforward to show that $\mathcal{L}_\diamond(e_{\mathcal{M},\bar{a}}) = \Sigma^*$ if and only if $\mathcal{M}$ does not accept on input $\bar{a}$. This finishes the proof of the EXPSPACE lower bound. □

**Proof of Proposition 3:**

We show explain how to adapt the reduction of Theorem 4 so that it does not longer uses parameterized expressions that are not simple.

Recall that the previous reduction used alphabet $\Delta = \{0, 1, \$, \%, \#, \&\} \cup \Gamma \cup (\Gamma \times Q)$.

For the simple case, we need a slightly bigger alphabet. Let $\Delta = \{0, 1, \$, \%, \#^e, \&^e, \#^o, \&^o\} \cup \Gamma \cup (\Gamma \times Q)$. The idea is to modify the way configurations are represented. Previously, we had that each configuration was represented by a word in

$$\#(\$[0]\%(\Gamma \cup (\Gamma \times Q))) \cdots (\$[2^n - 1]\%(\Gamma \cup (\Gamma \times Q)))\&.$$

In the modified reduction, however, configurations can be represented by either one of this expressions:

$$\#^e(\$[0]\%(\Gamma \cup (\Gamma \times Q))) \cdots (\$[2^n - 1]\%(\Gamma \cup (\Gamma \times Q)))\&^o, \text{ or}$$
$$\#^e(\$[0]\%(\Gamma \cup (\Gamma \times Q))) \cdots (\$[2^n - 1]\%(\Gamma \cup (\Gamma \times Q)))\&^o.$$

The intuition is that configurations using $\#^e$ and $\&^e$ represent an even stop of the computation of the Turing machine, whereas configurations using $\#^o$ and $\&^o$ represent an odd step.

It is now straightforward to modify expressions $e^{\text{form}}$, $e^i$ and $e^f$ to work under this codification of configurations: We just have to make sure that

- $e^{\text{form}}$ describes all the words that do not represent a concatenation of configurations of $\mathcal{M}$, where now valid configurations will have the form $\#^e u\&^e \#^o u'\&^o \#^e u''\&^e \cdots$, that is, valid configurations start in an even step, and then follow an even - odd - even - odd - ... pattern.

- $e^i$ describes words that do not start with the initial configuration of $\mathcal{M}$ over input $a$.

- $e^f$ describes words that do not end in a final configuration for $\mathcal{M}$

The descriptions of this expressions is omitted, since their extension is straightforward. Next, we show how one of the expressions in $e^{\text{trans}}$ is to be modified, the other remaining being analogous.

Consider expression $e_1^{\text{trans}}$, that intuitively accepts all words describing two configurations in which a cell not pointed by the head changed it's content. It was defined previously as

$$e_1^{\text{trans}} = \bigcup_{a \in \Gamma} \Delta^* \#(\$(0|1)^n \%(\Gamma \cup (\Gamma \times Q)))^* \$ x_1 \cdots x_n \% a (\$(0|1)^n \%(\Gamma \cup (\Gamma \times Q)))^* \&$$
$$\#(\$(0|1)^n \%(\Gamma \cup (\Gamma \times Q)))^* \$ x_1 \cdots x_n \%(\Gamma - \{a\} \cup (\Gamma \times Q) - (\{a\} \times Q))(\$(0|1)^n \%(\Gamma \cup (\Gamma \times Q)))^* \& \Delta^*$$

Then, we redefine it as $e_{1,e}^{\text{trans}} \mid e_{1,o}^{\text{trans}}$, where

$$e_{1,e}^{\text{trans}} = \bigcup_{a \in \Gamma} \Delta^* \#^e (\$(0|1)^n \%(\Gamma \cup (\Gamma \times Q)))^* \$$$
$$\left( x_1 \cdots x_n \big(\% a (\$(0|1)^n \%(\Gamma \cup (\Gamma \times Q)))^* \&^e \#^o (\$(0|1)^n \%(\Gamma \cup (\Gamma \times Q)))^* \$\big) \mid \right.$$
$$\left. \big(\%(\Gamma - \{a\} \cup (\Gamma \times Q) - (\{a\} \times Q))(\$(0|1)^n \%(\Gamma \cup (\Gamma \times Q)))^* \&^o\big) \right)^* (\#^e \Delta^* + \epsilon)$$

$$e_{1,o}^{\text{trans}} = \bigcup_{a \in \Gamma} \Delta^* \#^o (\$(0|1)^n \%(\Gamma \cup (\Gamma \times Q)))^* \$$$
$$\left( x_1 \cdots x_n \big(\% a (\$(0|1)^n \%(\Gamma \cup (\Gamma \times Q)))^* \&^o \#^e (\$(0|1)^n \%(\Gamma \cup (\Gamma \times Q)))^* \$\big) \mid \right.$$
$$\left. \big(\%(\Gamma - \{a\} \cup (\Gamma \times Q) - (\{a\} \times Q))(\$(0|1)^n \%(\Gamma \cup (\Gamma \times Q)))^* \&^e\big) \right)^* (\#^o \Delta^* + \epsilon)$$

And such that every appearance of $x_1, \ldots, x_n$ represents a new set of $n$ fresh variables. Notice then that these expressions are simple parameterized regular expressions. In order to see that the intended meaning of these expressions remains untouched, consider expressions

$$e_{1,e}'^{\text{trans}} = \bigcup_{a \in \Gamma} \Delta^* \#^e (\$(0|1)^n \%(\Gamma \cup (\Gamma \times Q)))^* \$ x_1 \cdots x_n \% a (\$(0|1)^n \%(\Gamma \cup (\Gamma \times Q)))^* \&^e$$
$$\#^o (\$(0|1)^n \%(\Gamma \cup (\Gamma \times Q)))^* \$ x_1 \cdots x_n \%(\Gamma - \{a\} \cup (\Gamma \times Q) - (\{a\} \times Q))(\$(0|1)^n \%(\Gamma \cup (\Gamma \times Q)))^* \&^o \Delta^*$$

$$e_{1,o}'^{\text{trans}} = \bigcup_{a \in \Gamma} \Delta^* \#^o (\$(0|1)^n \%(\Gamma \cup (\Gamma \times Q)))^* \$ x_1 \cdots x_n \% a (\$(0|1)^n \%(\Gamma \cup (\Gamma \times Q)))^* \&^o$$
$$\#^e (\$(0|1)^n \%(\Gamma \cup (\Gamma \times Q)))^* \$ x_1 \cdots x_n \%(\Gamma - \{a\} \cup (\Gamma \times Q) - (\{a\} \times Q))(\$(0|1)^n \%(\Gamma \cup (\Gamma \times Q)))^* \&^e \Delta^*$$

It is now easy to see that $\mathcal{L}(e_{1,e}'^{\text{trans}}) \subseteq \mathcal{L}(e_{1,e}^{\text{trans}})$ and $\mathcal{L}(e_{1,o}'^{\text{trans}}) \subseteq \mathcal{L}(e_{1,o}^{\text{trans}})$. Moreover, it is easy to check that none of the words in $\mathcal{L}(e_{1,e}^{\text{trans}})$ but not in $\mathcal{L}(e_{1,e}'^{\text{trans}})$ represent a valid sequence of configurations, and neither does any word in $\mathcal{L}(e_{1,o}^{\text{trans}})$ but not in $\mathcal{L}(e_{1,o}'^{\text{trans}})$. Using this observation, it is possible to modify all expressions in $\mathcal{L}(e^{\text{trans}})$ so that they are simple. We have omitted the rest of the proof since it goes along the same lines as the reduction of Theorem 4

## Proof of Proposition 4:

(Part 1) It is well known that CONTAINMENT$_\square$($e_1, \cdot$) is PSPACE-hard even for complete regular expressions. For the upper bound, let $e'_1$ be an expression such that $\mathcal{L}(e'_1) = \mathcal{L}_\square(e_1)$. Notice that, since $e_1$ is fixed and by Proposition 5, expression $e'_1$ can be computed in constant time. Then, it suffices to guess a valuation $\nu$ and a word $w$ such that $w \in \mathcal{L}(e'_1)$, but $w \notin \mathcal{L}(\nu(e_2))$, which can clearly be done in PSPACE.

(Part 2) We begin with the upper bound for the problem CONTAINMENT$_\square$($\cdot, e_2$). Assume that the input is a parameterized regular expression $e_1$, using variables in $\mathcal{W}$, and let $\Sigma$ be the alphabet of $e_1$. The CONP algorithm is as follows. First, construct a DFA $\mathcal{A}_{e_2}$ such that $\mathcal{L}(\mathcal{A}_{e_2}) = \mathcal{L}_\diamond(e_2)$, and then construct $\mathcal{A}^C_{e_2}$, the automaton that accepts the complement of $\mathcal{L}(\mathcal{A}_{e_2})$ (since $e_2$ is fixed, and by Proposition 5, this construction can be done in constant time). Next, guess a valuation $\nu$ from $\mathcal{W}$ to $\Sigma$, and, from $\nu(e_1)$, construct an automaton $\mathcal{A}_{\nu(e_1)}$ such that $\mathcal{A}_{\nu(e_1)} = \mathcal{L}(\nu(e_1))$ (It is a standard observation that this automaton can be constructed in polynomial time from $\nu(e_1)$). Finally, check that $\mathcal{A}_{\nu(e_1)} \cap \mathcal{A}^C_{e_2} \neq \emptyset$, which can be performed in polynomial time using a standard reachability test over the product of $\mathcal{A}_{\nu(e_1)}$ and $\mathcal{A}^C_{e_2}$. Let us show that this algorithm is sound and complete. If the intersection $\mathcal{A}_{\nu(e_1)} \cap \mathcal{A}^C_{e_2} \neq \emptyset$, then there is a word $w \in \mathcal{L}(\nu(e_1))$, and thus in $\mathcal{L}_\diamond(e_1)$, that does not belong to $\mathcal{L}_\diamond(e_2)$, or, in other words, that $\mathcal{L}_\diamond(e_1)$ is not contained in $\mathcal{L}_\diamond(e_2)$. On the other hand, it is clear that if $\mathcal{A}_{\nu(e_1)} \cap \mathcal{A}^C_{e_2} = \emptyset$ for all possible valuations $\nu$ from $\mathcal{W}$ to $\Sigma$, then $\mathcal{L}_\diamond(e_1)$ is contained in $\mathcal{L}_\diamond(e_2)$

The hardness is established via a reduction from 3-SAT to the complement of CONTAINMENT$_\square$($\cdot, e_2$). Let $e_2$ be the following regular expression over alphabet $\Sigma = \{0, 1, \#\}$:

$$e_2 = \big((10 \mid 01)^*\#((0 \mid 1)^3)^*000((0 \mid 1)^3)^*\big) \mid \big(((0 \mid 1)^2)^*(00 \mid 11)\Sigma^*\big) \mid \big(\Sigma^*\#\Sigma^*\#\Sigma^*\big),$$

and let $\varphi = \bigwedge_{1 \leq i \leq n}(\ell^1_i \vee \ell^2_i \vee \ell^3_i)$ be a propositional formula in 3-CNF over variables $\{p_1, \ldots, p_m\}$. That is, each literal $\ell^j_i$, for $1 \leq i \leq n$ and $1 \leq j \leq 3$, is either $p_k$ or $\neg p_k$, for $1 \leq k \leq m$. Next we show how to construct in polynomial time from $\varphi$ a parameterized regular expression $e_1$ over alphabet $\Sigma = \{0, 1, \#\}$ such that $\varphi$ is satisfiable if and only if $\mathcal{L}_\diamond(e_1) \not\subseteq \mathcal{L}(e_2)$.

Let $\mathcal{W} = \{x_i, \hat{x}_i \mid 1 \leq i \leq m\}$. Intuitively, each $x_i$ represents the value assigned to $p_i$, and $\hat{x}_i$ represents the value of $\neg p_i$. Moreover, assume that $h$ is a mapping from the literals $\ell^j_i$ ($1 \leq i \leq n$ and $1 \leq j \leq 3$) to $\mathcal{W}$, defined as expected: $h(\ell^j_i) = x_k$ if $\ell^j_i$ is $p_k$, for some $1 \leq k \leq m$, and $h(\ell^j_i) = \hat{x}_k$ if $\ell^j_i$ is $\neg p_k$.

Define $e_1$ as follows:

$$e_1 = x_1\hat{x}_1 \cdots x_m\hat{x}_m \# h(\ell^1_1)h(\ell^2_1)h(\ell^3_1) \cdots h(\ell^1_n)h(\ell^2_n)h(\ell^3_n).$$

We show that $\varphi$ is satisfiable if and only if $\mathcal{L}_\diamond(e_1) \not\subseteq \mathcal{L}(e_2)$.

($\Rightarrow$): Assume that $\varphi$ is satisfiable by valuation $\sigma$. Let $\nu$ be a valuation from $\mathcal{W}$ to $\Sigma$, defined as follows:

- For each $1 \leq k \leq m$, $\nu(x_k) = 1$ if $\sigma(p_k) = 1$, and $\nu(x_k) = 0$ otherwise.
- For each $1 \leq k \leq m$, $\nu(\hat{x}_k) = 0$ if $\sigma(p_k) = 1$, and $\nu(x_k) = 1$ otherwise.

Notice that $L(\nu(e_1))$ contains a single word. We shall abuse the notation and denote with $\nu(e_1)$ both this word and the aforementioned expression. It is clear that $\nu(e_1)$ contains a single symbol #, and starts with a prefix in $(01 \mid 10)^*\#$. Thus, if $\mathcal{L}_\diamond(e_1) \subseteq \mathcal{L}(e_2)$ it must be that $\nu(e_1)$ is denoted by the expression $(10 \mid 01)^*\#((0 \mid 1)^3)^*000((0 \mid 1)^3)^*$. But this implies that there are literals $\ell_i^1$, $\ell_i^2$ and $\ell_i^3$, for some $1 \leq i \leq n$, such that $\nu$ assigns the word 000 to $h(\ell_i^1)h(\ell_i^2)h(\ell_i^3)$. By construction of $\nu$, it must then be that $\sigma$ falsifies the $i$-th clause of $\varphi$, which contradicts the fact that $\sigma$ is a satisfying assignment.

($\Leftarrow$): Assume now that $\mathcal{L}_\diamond(e_1) \not\subseteq \mathcal{L}(e_2)$. By the definition of the $\diamond$-semantics, there is at least one valuation $\nu$ from $\mathcal{W}$ to $\Sigma$ such that $\mathcal{L}(\nu(e_1)) \not\subseteq \mathcal{L}(e_2)$. Notice again that, by the construction of $e_1$, $\nu(e_1)$ contains a single word. Again, we shall denote this word also by $\nu(e_1)$. Then if $\mathcal{L}(\nu(e_1)) \not\subseteq \mathcal{L}(e_2)$ it must be that $\nu(e_1)$ is not in $\mathcal{L}(e_2)$. This immediately entails that $\nu(e_1)$ cannot have two or more copies of the symbol $\#$, and thus we conclude that $\nu$ assigns to each variable $\mathcal{W}$ a symbol in $\{0, 1\}$. From the above observation, notice that the following valuation $\sigma$ for the variables in $\varphi$ is well defined:

- $\sigma(p_i) = 1$ if $\nu(x_i) = 1$, and $\sigma(p_i) = 0$ if $\nu(x_i) = 0$

Next, we show for all $1 \leq i \leq m$, it is the case that $\nu(x_i) \neq \nu(\hat{x}_i)$. Assume for the sake of contradiction that for some $1 \leq i \leq n$, we have that $\nu(x_i) = \nu(\hat{x}_i)$. From the construction of $e_1$, we then have that $\nu(e_1)$ is denoted by the expression $((0 \mid 1)^2)^*(00 \mid 11)\Sigma^*$, which contradicts the fact that $\nu(e_1)$ is not in $\mathcal{L}(e_2)$. Finally, we claim that $\varphi$ is satisfiable by valuation $\sigma$. Assume the contrary. Then there is a clause of form $(\ell_i^1 \vee \ell_i^2 \vee \ell_i^3)$, for some $1 \leq i \leq n$, such that, for each $1 \leq j \leq 3$, if $\ell_i^j$ is the literal $p_k$, for some $1 \leq k \leq m$, then $\sigma$ assigns the value 0 to $p_k$, and if $\ell_i^j$ is the literal $\neg p_k$, for some $1 \leq k \leq m$, then $\sigma$ assigns the value 1 to $p_k$. It is now straightforward to conclude that this fact contradicts the assumption that $\nu(e_1)$ is not in $\mathcal{L}(e_2)$, by studying all of the 8 possible cases.

### Proof of Proposition 5

Let $e$ be a parameterized regular expression over alphabet $\Sigma$, using variables in $\mathcal{W}$. Through this proof we heavily rely on the fact that there are $|\Sigma|^{|\mathcal{W}|}$ possible valuations $\nu : \mathcal{W} \to \Sigma$ for $e$.

First we show how to construct in double exponential time an NFA $\mathcal{A}_\square$ such that $\mathcal{L}_\square(e) = \mathcal{L}(\mathcal{A}_\square)$. For each valuation $\nu : \mathcal{W} \to \Sigma$, we denote by $\mathcal{A}_{nu(e)}$ the NFA such that $\mathcal{L}(\mathcal{A}_{\nu(e)}))\mathcal{L}(e)$ (This can be performed in PTIME by doing any standard regular expression to automata translation). Then, notice that $\mathcal{L}_\square(e) = \bigcap_{\nu:\mathcal{W}\to\Sigma} \mathcal{L}(\mathcal{A}_{\nu(e)})$, and thus we can just take the product of them. Given that there exists $|\Sigma|^{|\mathcal{W}|}$ possible valuations from $\mathcal{W}$ to $\Sigma$, and that each $\mathcal{A}_{\nu(e)}$ can be constructed in time $O(|e|\log^2 |e|)$ [13], the NFA for the product $\prod_{\nu:\mathcal{W}\to\Sigma} \mathcal{A}_{\nu(e)}$ can be constructed in time $|e|^{O(|\Sigma|^{|\mathcal{W}|})}$.

Next, in order to construct an automaton $\mathcal{A}_\diamond$ such that $\mathcal{L}_\diamond(e) = \mathcal{L}(\mathcal{A}_d m)$, one computes an NFA that represents $\bigcap_{\nu:\mathcal{W}\to\Sigma} \mathcal{A}_{\nu(e)}$ by combining all the automata with a nondeterministic choice at the beginning. Given that there exists $|\Sigma|^{|\mathcal{W}|}$ possible valuations from $\mathcal{W}$ to $\Sigma$, and that each $\mathcal{A}_{\nu(e)}$ can be constructed in time $O(|e|\log^2 |e|)$ [13], we have that the automaton $\mathcal{A}_\diamond$ can be constructed in time $|\Sigma|^{O(|\mathcal{W}|)} \cdot |e| \cdot \log^2 |e|$.

### Proof of Theorem 6

(Part 1) We begin with the double exponential bound for $\mathcal{L}_\square$. For each $n \in \mathbb{N}$, let $e_n$ be the following parameterized regular expression over alphabet $\Sigma = \{0, 1\}$ and variables $x_1, \ldots, x_n$:

$$e_n = ((0 \mid 1)^{n+1})^* \cdot x_1 \cdots x_n \cdot x_{n+1} \cdot ((0 \mid 1)^{n+1})^*.$$

Notice that each $e_n$ uses $n+1$ variables, and is of linear size in $n$. In order to show that every NFA deciding $\mathcal{L}_\Box(e_n)$ has $2^{2^n}$ states, we use the following result from [10]: if $L \subset \Sigma^*$ is a regular language, and there exists a set of pairs $P = \{(u_i, v_i) \mid 1 \le i \le m\} \subseteq \Sigma^* \times \Sigma^*$ such that

1. $u_i v_i \in L$, for every $1 \le i \le m$,

2. $u_j v_i \notin L$, for every $1 \le i, j \le m$ and $i \ne j$,

then every NFA accepting $L$ has at least $m$ states.

Given a collection $S$ of words over $\{0, 1\}$, let $w_S$ denote the concatenation, in lexicographical order, of all the words that belong to $S$, and let $w_{\bar{S},n}$ denote the concatenation of all words in $\{0,1\}^{n+1}$ that are not in $S$.

Then, define a set of pairs $P_n = \{(w_S, w_{\bar{S},n}) \mid S \subset \{0,1\}^{n+1}$ and $|S| = 2^n\}$. Since there are $2^{n+1}$ binary words of length $n+1$, the are $\binom{2^{n+1}}{2^n}$ different subsets of $\{0,1\}^{n+1}$ of size $2^n$, and thus $P_n$ contains $\binom{2^{n+1}}{2^n} \ge 2^{2^n}$ pairs.

Next, we show that $\mathcal{L}_\Box(e_n)$ and $P_n$ satisfy properties (1) and (2) above, which proves the double exponential lower bound.

1. We need to show that for every set $S \subset \{0,1\}^{n+1}$ of size $2^n$, the word $w_S, w_{\bar{S},n}$ belongs to $\mathcal{L}(\nu(e_n))$, for every possible valuation $\nu : \Sigma \to \{x_1, \dots x_{n+1}\}$. Let then $S$ be an arbitrary subset of $\{0,1\}^{n+1}$ of size $2^n$, and let $\nu$ be an arbitrary valuation from $\Sigma$ to $\{x_1, \dots, x_{n+1}\}$. Define $u = \nu(x_1) \cdots \nu(x_{n+1})$. Then $u$ is a substring of either $w_S$ or $w_{\bar{S},n}$. Assume the former is true (the other case is analogous). Then the word $w_S, w_{\bar{S},n}$ can be decomposed in $v \cdot u \cdot v' \cdot w_{\bar{S},n}$, with $v' \cdot v'' \in \mathcal{L}((0 \mid 1)^{n+1})$. This shows that $w_S, w_{\bar{S},n}$ belongs to $\mathcal{L}(\nu(e_n))$.

2. Assume for the sake of contradiction that there are distinct subsets $S_1, S_2$ of $\{0,1\}^{n+1}$ of size $2^n$ such that $w_{S_1} w_{\bar{S}_2,n}$ belongs to $\mathcal{L}_\Box(e_n)$. Since $S_1$ and $S_2$ are distinct, proper subsets of $\{0,1\}^{n+1}$ (they are of size $2^n$), there must be a word in $\{0,1\}^{n+1}$ that belongs to $S_2$ but not to $S_1$. Let $s$ be such word. Moreover, let $\nu$ be a valuation from $\Sigma$ to $\{x_1, \dots, x_{n+1}\}$ such that $\nu(x_1) \cdots \nu(x_{n+1}) = s$. It is straightforward to show the following:

    **Claim 3** *Let $u \in \{0,1\}^{n+1}$ be a word of size $n+1$. Then $u$ is a subword of every word $w \in \mathcal{L}_\Box(e_n)$. Moreover, there is a match for $u$ in $w$ that starts in a position $j$ of $w$ ($1 \le j \le |w|$), and such that $j = 1 \mod n+1$.*

    Since we have assumed that the word $w_{S_1} w_{\bar{S}_2,n}$ belongs to $\mathcal{L}_\Box(e_n)$, by the above claim we have that $s$ must be a subword of $w_{S_1} w_{\bar{S}_2,n}$ that matches $w_{S_1} w_{\bar{S}_2,n}$ in a position $j$ such that $j = 1 \mod n+1$. Then, from the construction of $w_{S_1}$ and $w_{\bar{S}_2,n}$, it must be that $s$ either belongs to $S_1$ or does not belong to $S_2$. This is a contradiction.

We use essentially the same technique to address the $\diamond$-semantics. To show the exponential lower bound for $\mathcal{L}_\diamond$, define $e_n = (x_1 \cdots x_n)^*$, and let $P_n = \{(w, w) \mid w \in \{0,1\}^n\}$. Clearly, $P_n$ contains $2^n$ pairs. All that is left to do is to show that $\mathcal{L}_\diamond(e_n)$ and $P_n$ satisfy properties (1) and (2) above.

1. From the fact that $\mathcal{L}_\diamond(e_n) = \bigcap_{w \in \{0,1\}^n} w^*$, we have that for each $u \in \{0,1\}^n$ the word $uu$ belongs to $\mathcal{L}_\diamond(e_n)$.

2. The same fact shows that for every $u, v \in \{0,1\}^n$, if $u \neq v$, then $uv \notin \bigcap_{w \in \{0,1\}^n} w^*$, and thus $uv \notin \mathcal{L}_\Diamond(e_n)$.

This finishes the proof of the theorem.

## Proof of Theorem 7:

It will be more convenient for us to work with automata than with regular expressions. We deal with NFAs with extended transitions, which can be not just of the form $(q, a, q')$, where $q$ and $q'$ are states, and $a \in \Sigma$, but also $(q, w, q')$, where $w \in \Sigma^*$. Such an automaton accepts a word $s \in \Sigma^*$ in the standard way: in a run, in state $q$, if the subword starting in the current position is $w$, it can skip $w$ and move to $q'$ if there is a transition $(q, w, q')$. Note that such automata are a mere syntactic convenience (they will appear as the results of applying valuations), as any such automaton $\mathcal{A}$ can be transformed, in polynomial time, into a usual NFA $\mathcal{A}'$ so that $\mathcal{L}(\mathcal{A}) = \mathcal{L}(A')$. Indeed, for each transition $t = (q, w, q')$ with $w = a_1 \ldots a_m$, introduce new states $q_t^1, \ldots, q_t^{m-1}$ and add transitions $(q, a_1, q_t^1), (q_t^1, a_2, q_t^2), \ldots, (q_t^{m-1}, a_n, q')$ to $\mathcal{A}'$. Thus, we shall work with automata with extended transitions.

Let $e$ be a parameterized regular expression with variables $x_1, \ldots, x_n$, whose domains are regular languages $L_1, \ldots, L_n$. Let $\mathcal{A}_e$ be an NFA equivalent to $e$, over the alphabet $\Sigma \cup \{x_1, \ldots, x_n\}$. If we have a valuation $\nu$ so that $\nu(x_i) \in L_i$ for each $i \leq n$, then $\nu(\mathcal{A}_e)$ is an automaton with extended transitions: in it, each transition $(q, x_i, q')$ is replaced by $(q, \nu(x_i), q')$. It is then immediate from the construction that $\mathcal{L}(\nu(e)) = \mathcal{L}(\nu(\mathcal{A}_e))$ and thus $\mathcal{L}_\Box(e) = \bigcap_\nu \mathcal{L}(\nu(\mathcal{A}_e))$.

Next, consider *finitary* valuations $\nu$, which are partial functions defined on variables $x_i$ such that $L_i$ is a finite languag of course $\nu(x_i) \in L_i$. On variables $x_j$ with infinite $L_j$ such valuations are undefined. By $\nu(\mathcal{A}_e)$ we mean the automaton (with extended transition) resulting from $\mathcal{A}_e$ as follows. First, all transitions $(q, a, q')$, where $a$ is a letter, are kept. Second, if $(q, x_i, q')$ is a transition, then $\nu(\mathcal{A}_e)$ contains $(q, \nu(x_i), q')$ only if $\nu(x_i)$ is defined. In other words, transitions using variables whose domains are infinite, are dropped.

Let $\nu_1, \ldots, \nu_M$ enumerate all the finitary valuations (clearly there are finitely many of them). Let $\mathcal{A}_i = \nu_i(\mathcal{A}_e)$, for $i \leq M$. We now show that $\mathcal{L}_\Box(\mathcal{A}_e) = \bigcap_{i \leq M} \mathcal{L}(\mathcal{A}_i)$.

First, if $\nu_i$ is a finitary valuation and $\nu$ is any extension of $\nu_i$ to a valuation on all the variables $x_1, \ldots, x_n$, then clearly $\mathcal{L}(\nu_i(\mathcal{A}_e)) \subseteq \mathcal{L}(\nu(\mathcal{A}_e))$. Note that every valuation $\nu$ is an extension of some finitary valuation $\nu_i$, and thus $\mathcal{L}_\Box(\mathcal{A}_e) = \bigcap_{\text{all valuations } \nu} \mathcal{L}(\nu(\mathcal{A}_e)) \supseteq \bigcap_{i \leq M} \mathcal{L}(\nu_i(\mathcal{A}_e))$. For the reverse inclusion, let $w \in \mathcal{L}_\Box(\mathcal{A}_e)$; in particular, $w \in \mathcal{L}(\nu(\mathcal{A}_e))$ for every valuation $\nu$. Take an arbitrary finitary valuation $\nu_i$ and let $V_i$ be the (infinite) set of all the valuations $\nu$ that extend $\nu_i$. Let $V_i(w)$ be the subset of $V_i$ that contains valuations $\nu$ with the property that for each variable $x_j$ with an infinite domain $L_j$, we have $|\nu(x_j)| > |w|$; clearly $V_i(w)$ is an infinite set as well. Take any $\nu \in V_i(w)$; we know from $w \in \mathcal{L}_\Box(\mathcal{A}_e)$ that $w \in \mathcal{L}(\nu(\mathcal{A}_e))$. In particular, there is an accepting run of $\nu(\mathcal{A}_e)$ that never uses any transition $(q, \nu(x_j), q')$ with $L_j$ infinite, since $|\nu(x_j)| > |w|$. Thus, such an accepting run may only use transitions resulting from valuations of variables with finite domains, and hence it is also an accepting run of $\nu_i(\mathcal{A}_e)$. This shows that $w \in \mathcal{L}(\nu_i(\mathcal{A}_i))$; since $\nu_i$ was chosen arbitrarily, it means that $w \in \bigcap_{i \leq M} \mathcal{L}(\nu_i(\mathcal{A}_e))$, and thus proves $\mathcal{L}_\Box(\mathcal{A}_e) = \bigcap_{i \leq M} \mathcal{L}(\nu_i(\mathcal{A}_e)) = \bigcap_{i \leq M} \mathcal{L}(\mathcal{A}_i)$.

This immediately shows that $\mathcal{L}_\Box(e) = \mathcal{L}_\Box(\mathcal{A}_e)$ is regular, as a finite intersection of regular languages. Lower bounds on complexity apply immediately as they were all established for the case when each $L_i = \Sigma$. So we need to prove upper bounds. To do so, one can see, by analyzing the proofs for the case when all domains are $\Sigma$, that it suffices to establish the following facts on the set of automata $\mathcal{A}_i$, for $i < M$:

- $M$ is at most exponential in the size of the input;
- checking whether a given automaton $\mathcal{A}$ is one of the $\mathcal{A}_i$'s can be done in time polynomial in the size of $\mathcal{A}$; and
- for each $\mathcal{A}_i$, for $i < M$, its size it at most polynomial in the size of the input.

(To give a couple of examples, to see the EXPSPACE-bound on NONEMPTINESS$_\square$, we construct exponentially many automata of polynomial size and check nonemptiness of their intersection. To see the NP upper bound on MEMBERSHIP$_\diamond$, one guesses a polynomial-size $\mathcal{A}_i$, checks in polynomial time that it is indeed a correct automaton, and then checks again in polynomial time whether a given word is accepted by it.)

Recall that the input to the problem we are considering is $(e; \bar{L})$, or $(\mathcal{A}_e; \bar{L})$, and we can assume that each $L_i$ is given by an NFA $\mathcal{B}_i$ (if part of the input is a regular expression, we can convert it into an NFA in $O(n \log^2 n)$ time [13]).

To show the bounds, assume without loss of generality that from each $\mathcal{B}_i$ all nonreachable states, and states from which final states cannot be reached, are removed (this can be done in polynomial time). Then $\mathcal{L}(\mathcal{B}_i)$ is finite iff $\mathcal{B}_i$ does not have cycles. Thus, if $n_i$ is the number of states of $\mathcal{B}_i$, then the longest word accepted by $\mathcal{B}_i$ is of length $n_i$, and hence the size of each finite $L_i = \mathcal{L}(\mathcal{B}_i)$ is at most $|\Sigma|^{n_i+1}$. Hence, the total number of all the words in finite languages $L_i$'s is less than $|\Sigma|^N$, where $N = n + \sum n_i$, with the sum taken over indexes $i$ such that $L_i$ is finite. This means that in turn the number of finitary valuations $M$, i.e. mappings from some of the variables $x_i$'s into words in these finite languages is at most $|\Sigma|^{Nn}$, which is thus exponential in the size of the input.

The remaining two properties are easy. Since the length of each word accepted by one of the $\mathcal{B}_i$'s is at most the number of states in $\mathcal{B}_i$, the size of all the automata $\nu_i(\mathcal{A}_e)$ is bounded by a polynomial in the size of the input; changing extended transitions in those to the usual NFA transitions involves only a linear increase of size. To check whether an automaton $\mathcal{A}$ is one of the $\mathcal{A}_i$'s, we check whether all its transitions involving both states from $\mathcal{A}_e$ come from $\mathcal{A}_e$ or from a single-letter valuation. Every other transition must be on a path between two states from $\mathcal{A}_e$. One reads words on these paths, and checks if they form a finitary valuation. This can easily be done in polynomial time.